\documentclass[5p]{elsarticle}

\usepackage{hyperref}
\usepackage{booktabs} 
\usepackage{amsmath,amssymb,amsfonts}
\usepackage{enumitem}
\usepackage{graphicx}
\usepackage{xcolor}
\usepackage{url}
\usepackage{multirow}
\usepackage{xspace}
\usepackage{subcaption}
\usepackage{balance}
\usepackage{pgf}
\usepackage{hhline}
\usepackage{pifont}
\usepackage[capitalize, noabbrev]{cleveref}
\usepackage[most]{tcolorbox}

\journal{Journal of \LaTeX\ Templates}

\definecolor{myblue}{RGB}{0, 43, 136}
\definecolor{plus}{RGB}{0,163,0}
\definecolor{minus}{RGB}{163,0,0}
\newcommand{\argmax}{\mathop{\mathrm{argmax}}}   
\newcommand{\name}{SIMFL\xspace}
\newcommand{\se}{Seshat\xspace}
\newcommand{\dfj}{\textsc{Defects4J}\xspace}

\newcommand{\cmark}{\ding{51}}%
\newcommand{\xmark}{\ding{55}}%
\newcommand{\fix}[1]{\textcolor{black}{#1}}

\begin{document}

\begin{frontmatter}

\title{Learning Test-Mutant Relationship for Accurate Fault Localisation}


\author[kaist]{Jinhan Kim}
\author[kaist]{Gabin An}
\author[chalmers]{Robert Feldt}
\author[kaist]{Shin Yoo\corref{mycorrespondingauthor}}

\address[kaist]{School of Computing, KAIST, Daejeon, Republic of Korea}
\address[chalmers]{Dept. of Computer Science and Engineering, Chalmers University of Technology, Gothenburg, Sweden}
\cortext[mycorrespondingauthor]{Corresponding author}
\ead{shin.yoo@kaist.ac.kr}

\begin{abstract}
  \textbf{Context}: Automated fault localisation aims to assist
  developers in the task of identifying the root cause of the fault by narrowing
  down the space of likely fault locations. Simulating variants of the faulty
  program called mutants, several Mutation Based Fault Localisation (MBFL) 
  techniques have been proposed to automatically locate faults. Despite their 
  success, existing MBFL techniques suffer from the cost of performing mutation 
  analysis after the fault is observed.

  \noindent \textbf{Method}: To overcome this shortcoming, we propose a
  new MBFL technique named \name (Statistical Inference for Mutation-based
  Fault Localisation). \name localises faults based on the past results of 
  mutation analysis that has been done on the earlier version in the project 
  history, allowing developers to make predictions on the location of incoming 
  faults in a just-in-time manner. Using several statistical inference 
  methods, \name models the relationship between test results of the mutants 
  and their locations, and subsequently infers the location of the current faults.

  \noindent \textbf{Results}: The empirical study on \dfj dataset shows
  that \name can localise 113 faults on the first rank out of 224 faults, outperforming
  other MBFL techniques. Even when \name is trained on the predicted kill
  matrix, \name can still localise 95 faults on the first rank out of 194 faults.
  Moreover, removing redundant mutants significantly improves the localisation
  accuracy of \name by the number of faults localised at the first rank up to 51.

  \noindent \textbf{Conclusion}: This paper proposes a new MBFL technique
  called \name, which exploits ahead-of-time mutation analysis to localise
  current faults. \name is not only cost-effective, as it does not need a
  mutation analyse after the fault is observed, but also capable of localising
  faults accurately.
\end{abstract}


\end{frontmatter}

\section{Introduction}
\label{sec:intro}
Fault Localisation (FL) is the problem of diagnosing the root cause of the
fault by highlighting a few program elements likely to be responsible for the fault. It is an
expensive debugging activity as it involves a human inspection to point out the
suspicious locations in Program Under Test (PUT) based on the understanding of
the root cause. As such, the necessity of developing automated localisation
techniques has received increasing
attention~\cite{Yoo:2012kx,Sohn:2017xq,Abreu:2011kq,Naish:2011fk,Yoo:2017ss,Wong:2007rt},
not only to help human developers but also to be adopted as a prerequisite for
Automated Program Repair (APR)~\cite{Qi:2013fk,Weimer:2009fk,Wen2018dk}.

 Mutation Based Fault Localisation
  (MBFL)~\cite{Moon:2014ly,Hong:2017qy,Papadakis:2012fk,Papadakis:2015sf,Pearson:2017aa}
  is one such way that utilises mutation analysis. Mutants are program
  variants made up of a single or multiple syntactic operator(s), designed to
  simulate the real faults. Various empirical studies~\cite{off92, Just2014fse,
  chen2020revisiting} have provided evidence of a correlation between detection
  of the mutants and detection of the faults. Exploiting this,
  Metallaxis~\cite{Papadakis:2012fk,Papadakis:2015sf} considers the location of
  the mutants that have similar failing patterns with the faults as candidate
  locations of the fault. MUSE~\cite{Moon:2014ly} and MUSEUM~\cite{Hong:2017qy}
  have taken different approaches that reward the mutants causing partial fixes
  and devalue the mutants making an additional failure.

Although the usefulness of MBFL techniques has long been demonstrated,
  the substantial cost of mutation analysis has hampered their practical
  applications. The cost reduction techniques for MBFL are in line with the cost
  reduction techniques for traditional mutation testing: random sampling of the
  mutants~\cite{Papadakis:2015sf} and higher-order
  mutants~\cite{jang2021hotfuz}. They have reduced the number of mutants and
  test executions, but still they are not suitable for large systems like
  Google's on which more than ten million test executions take place
  everyday~\cite{petrovic2021does}. To avoid this, recent advances employ the
  context of Continuous Integration (CI)~\cite{meyer2014continuous}: predictive
  mutation testing techniques~\cite{zhang2018predictive, mao2019extensive,
  Kim2021ax} aim to predict mutation testing results without mutant executions
  by learning static and dynamic features of mutants from earlier version of the
  project.

 In this work, we propose \name (Statistical Inference for Mutation-based
  Fault Localisation), an MBFL technique that operates in the CI context. Unlike
  other FL techniques that undertake work on locating faults \emph{a
  posteriori}, \name uses a kill matrix computed on earlier version of the PUT,
  which therefore amortises the costs of mutation analysis that has been done
  after the actual faults are observed. The kill matrix records the test results
  of each mutant, and \name uses it to relate the locations of mutants and
  faults: if the mutants were killed by the similar set of tests that detects the
  faults, \name values such mutants and presumes their locations as likely
  locations of the faults. To model this kind of predictive inferences, \name
  makes use of several statistical inference techniques such as Bayesian
  inference, probabilistic coupling, and more sophisticated models such as
  Logistic Regression or Multi-Layer Perceptron. Once building a predictive
  model in advance, \name produces a rank of suspicious program elements by
  feeding the test results of the fault into the model.

 We conduct extensive empirical experiments on \name in three different use
  cases. Using the real-world fault dataset, \dfj v1.3.1~\cite{Just:2014aa}, and
  Major mutation tool~\cite{Just2011gq}, \name places 113 faults on the first rank out
  of 224 faults and outperformed existing MBFL techniques. Even if \name is
  built on the predicted kill matrix by \se~\cite{Kim2021ax}, it localises 95
  faults on the first rank out of 194 faults, showing its practical value. As
  \name does not require the execution of mutants, it shows a significant
  improvement in efficiency over other MBFL techniques and it is even faster
  than running a entire test suite once. We also investigate how much \name is
  affected by mutant sampling and the results show that \name retains 80\% of
  localisation accuracy when it uses only 10\% of mutants. Lastly, we
  demonstrate the assumption that subsumed mutants~\cite{just2012redundant,
  ammann2014establishing} would disrupt the scoring functions of \name. The
  evaluation of \name models after eliminating the subsumed mutants reveals that
  the number of faults localised at the first rank increases up to 51.

  This paper is an extended version of the conference paper~\cite{Kim:9700328}.
  The contributions of the conference paper are as follows:
  \begin{itemize}
    \item We present \name, an MBFL technique that infers the location of faults
    based on the mutation analysis performed before the faults are observed.
    This allows \name to significantly amortise the costs of mutation analysis,
    and the experiments on \dfj dataset showed that \name achieves better
    localisation accuracy than existing MBFL techniques.
    \item We introduce several modelling schemes of \name using statistical
    inferences and machine learning techniques. We not only compare their
    effectiveness but also provide the empirical evidence of the impacts of
    model viability and mutant sampling.
    \item We further investigate the impact of test granularity and mutant
    filtering based on the kill reason (e.g., exception). The in-depth
    observation on the relationship with other FL techniques open up a positive
    hybridisation of \name and other techniques.
  \end{itemize}

  The paper has been extended with the following technical contributions:
  \begin{itemize}
    \item We have added a new use scenario of \name using a predicted kill
    matrix of \se, which is not dependent on pre-existence of failing tests.    
    \item We have proposed a new ranking model based on probabilistic coupling.
    We apply a definition of mutant-fault coupling that values the mutants tightly
    coupled to the fault.
    \item We have conducted a new experiment to investigate the execution time
    of \name compared to other MBFL techniques.
    \item We have added new faults introduced in \dfj v2.0.0 in our study, to
    further validate our conclusion with a larger fault dataset.
    \item We have investigated the effects of removing the subsumed mutants. We
    observed that the subsumed mutants disturb and inflate the method-level
    aggregation of FL scores by \name. The new experiment showed that
    eliminating them significantly improves the effectiveness of \name.
    \item We have made a raw data publicly available at
    \url{https://figshare.com/s/954a7f27bb0e08b009f8} for the replication of
    our experiments.
  \end{itemize}

 The rest of the paper is organised as follows.
  Section~\ref{sec:methodology} details how the mutation analysis results are
  formulated for fault localisation. Section~\ref{sec:exp_design} introduces
  three use scenarios of \name, research questions, and experimental settings.
  Section~\ref{sec:result} answers each research question and
  Section~\ref{sec:discussion} provides discussions. Section~\ref{sec:threats}
  presents threats to validity. Section~\ref{sec:related_work} introduces
  related work and Section~\ref{sec:conclusion} concludes.

\section{Methodology}
\label{sec:methodology}

Intuitively, the underlying assumption of \name is that,
for a test that has killed the mutants located on a specific program 
element, the same program element should be identified as the suspicious 
location when the same test later fails again. This is based on the coupling effect 
hypothesis in mutation testing: essentially we \emph{simulate} the occurrence
of real faults with artificial faults with known locations, i.e., mutants,
and build predictive models for actual future faults. This section describes
the models and the statistical inference techniques used by \name.

\begin{figure}[ht]
  \centering  
  \includegraphics[width=.5\textwidth]{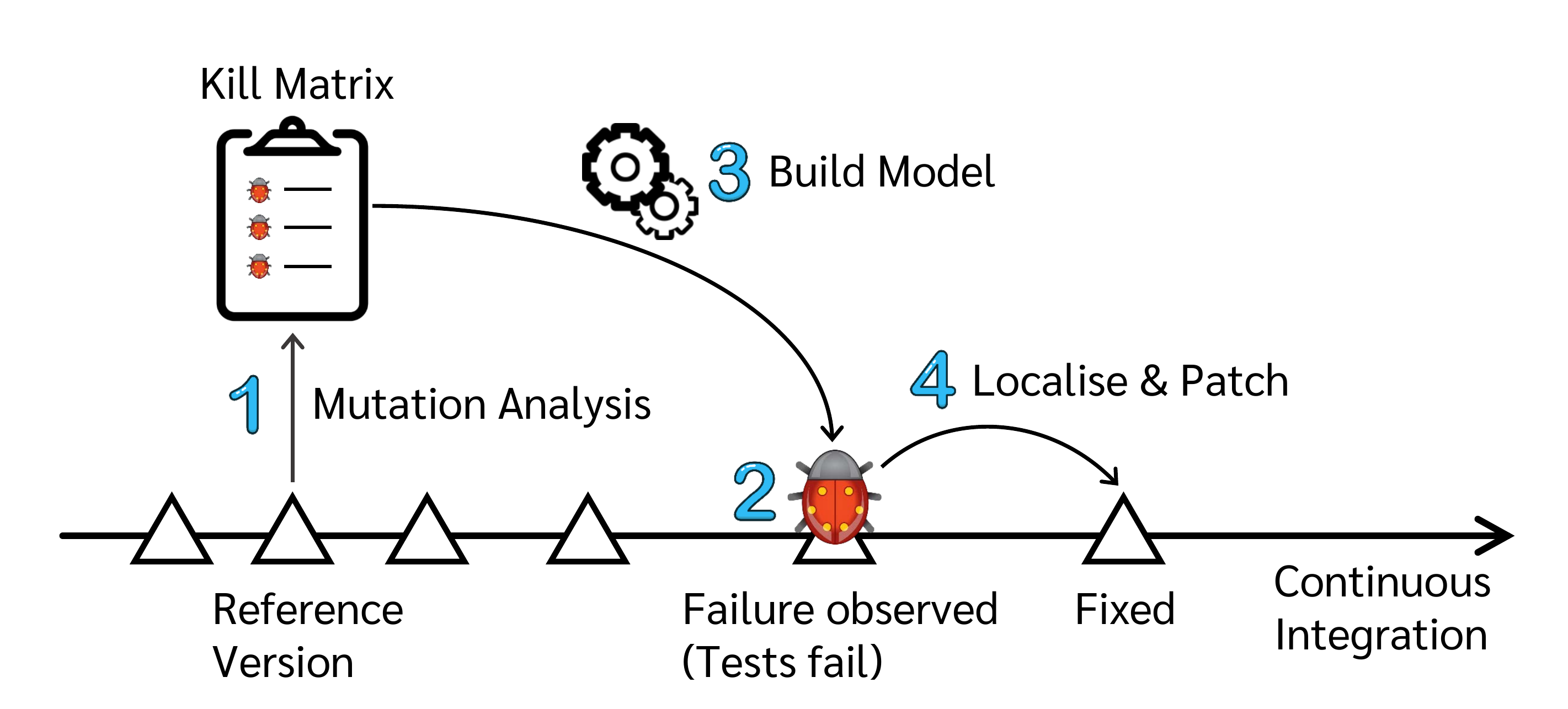}  
  \caption{Expected use case scenario of \name \label{fig:scenario}}  
\end{figure}

Figure~\ref{fig:scenario} depicts the expected use case scenario of \name, 
which includes four stages:

\begin{enumerate}
\item Perform mutation analysis for a version of SUT, and produce the kill matrix. The version is called the \emph{reference} version.
\item While testing a subsequent version, a failure is observed.
\item Using the information of which test case(s) failed, as well as the kill matrix, build a predictive model for fault localisation.
\item Guided by the localisation result, patch the fault.
\end{enumerate}

\begin{table}[ht]
  \centering
  \caption{Key Symbols \label{tab:key_symbols}}
  \scalebox{0.7}{
    \begin{tabular}{l|l}
    \toprule
    Symbol & Meaning\\
    \midrule
    $\mathbf{P}$ & Program under test. \\
    $\mathbf{T}$ & Entire test suite. \\
    $\mathbf{T}_f$ & Set of failing tests on a given program. \\
    $\mathbf{T}_p$ & Set of passing tests on a given program. \\
    $\mathbf{K}$ & Kill matrix. \\
    $m$ & Mutant. \\
    $e$ & Program element. \\
    $\mathbf{K}_m$ & Set of tests that kill $m$. \\
    $\mathbf{X}_e$ & Set of mutants located on $e$. \\
    $M_e$ & Event that $e$ is mutated. \\
    $F_t$ & Event that a test case $t$ fails on a given program. \\
    $P_t$ & Event that a test case $t$ passes on a given program. \\
    $F_{\mathbf{T}_f}$ & Event that all tests in $\mathbf{T}_f$ fail on a given program. \\
    $Pr\left(M_e\right)$ & Probability of $e$ is mutated. \\
    $\fix{Pr}\left(F_t \mid M_e \right)$ & Probability of $t$ fails when $e$ is mutated. \\
    \bottomrule
    \end{tabular}  
  }  
\end{table}

\subsection{Mutation Analysis}
\label{sec:mutation_analysis}

\begin{table*}[ht]
  \centering
  \caption{An Example Kill Matrix \label{tab:kill_matrix}}
  \scalebox{1.0}{
  \begin{tabular}{l|l|l|rrrr|rrrr}
    \toprule
    Class & Method & Mutants & \multicolumn{4}{c|}{Test Results} & \multicolumn{4}{c}{0-1 Vectors} \\    
    & & & $t_1$ & $t_2$ & $t_3$ & $t_4$ & \multicolumn{4}{c}{of $t_1, t_2, t_3, t_4$} \\
    \midrule
    \multirow{5}{*}{\shortstack[l]{com.\\acme.\\Foo}} & \multirow{3}{*}{getType} & n.isName() $\mapsto$ true ($m_1$) & \xmark& \xmark& \cmark& \cmark &\multicolumn{4}{c}{$\left(1, 1, 0, 0\right)$}\\ 
    & & n.isName() $\mapsto$ false ($m_2$) & \cmark& \cmark& \xmark& \cmark & \multicolumn{4}{c}{$\left(0, 0, 1, 0\right)$}\\
    & & bFlag or isInferred $\mapsto$ isInferred ($m_3$) & \cmark& \xmark& \xmark& \cmark &\multicolumn{4}{c}{$\left(0, 1, 1, 0\right)$}\\    
    \cmidrule{2-11}
    & \multirow{2}{*}{\shortstack[l]{resolveType}} & param.isTemplateType() $\mapsto$ true ($m_5$) & \cmark& \xmark& \cmark& \cmark &\multicolumn{4}{c}{$\left(0, 1, 0, 0\right)$}\\ 
    & & resolvedType() $\mapsto$ deleted ($m_6$) & \xmark& \xmark& \xmark& \cmark &\multicolumn{4}{c}{$\left(1, 1, 1, 0\right)$}\\    
    \bottomrule
  \end{tabular}}
\end{table*}

We perform mutation analysis on the reference version of a program $\mathbf{P}$
with a test suite $\mathbf{T}$, and compute a kill matrix $\mathbf{K}$, which
contains a complete report of all tests executed on all mutants. An example kill
matrix is shown in Table~\ref{tab:kill_matrix}: the mutant $m_1$ located in the
method \texttt{getType} is killed by the test cases $t_1$ and $t_2$, whereas
$m_2$ is only killed by $t_3$. We list the key symbols in
Table~\ref{tab:key_symbols} to aid in the understanding of the formulae we will
introduce in the following.

Let $\mathbf{K}_m$ denote a set of tests that kill mutant $m$,
let $\mathbf{X}_e$ be a set of mutants located on a program element $e \in 
\mathbf{P}$,
let $M_e$ be an event that $e$ is mutated, and let $F_t$ be an event that a 
test case $t$ fails on a given program.
Based on the kill matrix $\mathbf{K}_m$, we can approximate the probability of
test case $t$ killing the mutants located on the program element $e$ as follows:

\small
\begin{equation}
  \fix{Pr}\left(F_t \mid M_e \right) \simeq
	\frac{|\left\{m \in \mathbf{X}_e : t \in \mathbf{K}_m\right\}|}{|\mathbf{X}_e|}
\label{eq:prob_org}
\end{equation}
\normalsize

Note that this is strictly an \emph{approximation} based on the observed kill
matrix because it is impossible to produce and evaluate all possible mutants.
The value of $Pr\left(M_e\right)$ is the ratio of the number of all possible
mutants on $e$ to the number of all possible mutants on $\mathbf{P}$; for
$\fix{Pr}\left(F_t \mid M_e \right)$, we need to calculate the number of all
possible mutants in $e$ that are killed by $t$. Neither is feasible.
Consequently, we assume that we can analyse a \emph{finite} set of mutants that
allow us to approximate Equation~\ref{eq:prob_org}. We note that test failure history over the project lifetime can be used to approximate both $P(M_e)$ (i.e., the probability that method $e$ is faulty after a commit) and $P(F_t|M_e)$ (i.e., the probability that test $t$ fails when method $e$ is faulty). Such historical approximation may be more accurate, but would require high traceability between test failures and code changes throughout the project history.

Next, using Bayes' rule, we calculate the revised probability of the event that 
the program element $e$ has been mutated, given that the test case $t$ fails:

\small
\begin{equation}
  \begin{split}
    \fix{Pr}\left(M_e \mid F_t \right)
    &= \frac{\fix{Pr}\left(F_t \mid M_e \right)\fix{Pr}\left(M_e\right)}{\fix{Pr}\left(F_t\right)} \\
    &\simeq \fix{Pr}\left(\text{fault exists in } e \mid F_t\right)
  \end{split}
\label{eq:prob_bayes}
\end{equation}
\normalsize

We argue that, if real faults are coupled to mutants, the probability above 
can approximate the likelihood that the fault is located on the program 
element $e$, when $t$ is a failing test case in the future. Our 
approximation assumes that tests are equally sensitive to mutants 
and real bugs. This allows us to make ranking models that sort the program 
elements in descending order of the probability.

\subsection{Ranking Models Based on Bayes' Rule}
\label{sec:ranking_model_bayes_rule}

We regard the probability in Equation~\ref{eq:prob_bayes} as the quantitative 
score representing how suspicious the program element $e$ is for the failure observed via
the failure of $t$. This section presents the formulations of ranking models 
based on the scores as well as more refined inference models based on kill matrix data.

\subsubsection{Exact Matching (EM)}
\label{sec:exact_matching}

This model is an extension of Equation~\ref{eq:prob_bayes} to a set of test 
cases. Let $\mathbf{T} = \{t_i : 1 \leq i \leq n \leq n'\}$ be the test set,
which consists of two disjoint sets: $\mathbf{T}_f = \{t_1, \dots, t_n\}$ is 
the set of failing test cases, and $\mathbf{T}_p = \mathbf{T} \setminus \mathbf
{T}_f$ is the set of passing tests, on the faulty program. While there can be
many different formulations of ranking models based on a set of test cases, we
start by treating the set of all observed failures, $F_{\mathbf{T}_f}$, as
a conjunctive event of individual test case failures, 
i.e., $F_{\mathbf{T}_f} = F_{t_1} \cap \dots \cap F_{t_n}$. Our goal is to 
find the faulty program element $e_i \in \mathbf{P}$ with the highest probability of being the cause of the observed failure symptoms, that is, 
$\fix{Pr}\left(M_{e_i} \mid F_{\mathbf{T}_f}\right)$. It follows that:

\small
\begin{equation}
  \begin{split}
    \argmax_{i}\fix{Pr}\left(M_e \mid F_{\mathbf{T}_f}\right)
     &= \argmax_{i}\frac{\fix{Pr}\left(F_{\mathbf{T}_f} \mid M_e\right)\fix{Pr}\left(M_e\right)}{\fix{Pr}\left(F_{\mathbf{T}_f}\right)}
  \end{split}  
  \label{eq:prob_bayes_test_set}
\end{equation}
\normalsize

The denominator in Equation~\ref{eq:prob_bayes_test_set}, $\fix{Pr}\left(F_{\mathbf{T}_f}
\right)$, can be ignored without affecting the order of ranking based on this 
score, because it is not related to a specific program element. Expanding the 
numerator yields the following:
\small
\begin{equation}
  \begin{split}
    &\argmax_{i}\fix{Pr}\left(F_{\mathbf{T}_f} \mid M_{e_i}\right)\fix{Pr}\left(M_{e_i}\right) \\
    &= \argmax_{i}\fix{Pr}\left(F_{t_1} \cap \dots \cap F_{t_n} \mid M_{e_i}\right)\fix{Pr}\left(M_{e_i}\right) \\
    &= \argmax_{i}\frac{|\{m \in \mathbf{X}_{e_i} : \{t_1, \dots, t_n\} = \mathbf{K}_m\}|}{|\mathbf{X}_{e_i}|}\fix{Pr}\left(M_{e_i}\right) \\
    &= \argmax_{i}\frac{|\{m \in \mathbf{X}_{e_i} : \mathbf{T}_f = \mathbf{K}_m\}|}{|\mathbf{X}_{e_i}|}\frac{|\mathbf{X}_{e_i}|}{|\mathbf{X}_{\mathbf{P}}|} \\
    &= \argmax_{i}\frac{|\{m \in \mathbf{X}_{e_i} : \mathbf{T}_f = \mathbf{K}_m\}|}{|\mathbf{X}_{\mathbf{P}}|} \\
    &= \argmax_{i}|\{m \in \mathbf{X}_{e_i} : \mathbf{T}_f = \mathbf{K}_m\}| 
  \end{split}  
  \label{eq:argmax_only_failing}
\end{equation}
\normalsize

Intuitively, Equation~\ref{eq:argmax_only_failing} counts the mutants on $e$
that cause the same set of test cases to fail as the symptom of the actual 
fault, $F_{\mathbf{T}_f}$. We call this model the Exact Matching (EM) model 
with failing test cases, denoted by EM(F).

Alternatively, we can include passing tests in the pattern matching as well.
Let $P_t$ be an event that a test case $t$ passes on a given program, then
Equation~\ref{eq:argmax_only_failing} changes as follows:

\small
\begin{equation}
  \begin{split}
    &\argmax_{i}\fix{Pr}\left(F_{\mathbf{T}_f} \cap P_{\mathbf{T}_p} \mid M_{e_i}\right)\fix{Pr}\left(M_{e_i}\right) \\
    &= \argmax_{i}\fix{Pr}\left(F_{t_1} \cap \dots \cap F_{t_n} \cap P_{t_{n+1}} \cap \dots \cap P_{t_{n'}} \mid M_{e_i}\right)\fix{Pr}\left(M_{e_i}\right) \\
    &= \argmax_{i}|\{m \in \mathbf{X}_{e_i} : \mathbf{T}_f = \mathbf{K}_m \wedge \mathbf{T}_p = \mathbf{T} \setminus \mathbf{K}_m \}|
  \end{split}  
  \label{eq:argmax_with_passing}
\end{equation}
\normalsize

Similarly to EM(F), this model is called EM(F+P): it counts the mutants on $e$
that cause the same set of test cases to fail and pass exactly as the symptom 
of the actual fault. If, for example, a test case $t$ passed under the actual fault, EM(F+P) model will not count any mutants that are killed by $t$.

\subsubsection{Partial Matching (PM)}
\label{sec:partial_matching}
The Exact Matching (EM) models lose any partial matches between the symptom 
and the mutation results. Suppose two test cases, $t_1$ and $t_2$, failed 
under the actual fault, but only $t_1$ killed a mutant on the faulty program 
element, i.e., $ \exists t_1, t_2 \in \mathbf{T}_f, t_1 \in \mathbf{K}_m 
\land t_2 \notin \mathbf{K}_m $. The information that $t_1$ kills a mutant on the location of the fault is lost, simply because $t_2$ failed to do the same. To retrieve this partial information, we propose two additional models based on partial matches: a multiplicative partial match model and an additive partial match model.

\begin{itemize}
    \item PM$^*$(F): Multiplicative Partial Match Model w/ Failing Tests

    \small
    \begin{equation}
        \begin{split}
          &\argmax_{i} \prod_{t \in \mathbf{T}_f} \left(\fix{Pr}\left(M_{e_i} \mid F_{t}\right) + \epsilon\right)\\
          &=\argmax_{i} \prod_{t \in \mathbf{T}_f} \frac{\fix{Pr}\left(F_{t}\mid M_{e_i}\right)\fix{Pr}\left(M_{e_i}\right)}{\fix{Pr}\left(F_t\right)} + \epsilon\\
          &=\argmax_{i} \prod_{t \in \mathbf{T}_f} \fix{Pr}\left(F_{t}\mid M_{e_i}\right)\fix{Pr}\left(M_{e_i}\right) + \epsilon\\
          &=\argmax_{i} \prod_{t \in \mathbf{T}_f} \frac{|\{m \in \mathbf{X}_{e_i} : t \in \mathbf{K}_m\}|}{|\mathbf{X}_{e_i}|}\frac{|\mathbf{X}_{e_i}|}{|\mathbf{X}_{\mathbf{P}}|} + \epsilon\\
          &=\argmax_{i} \prod_{t \in \mathbf{T}_f} \left(|\{m \in \mathbf{X}_{e_i} : t \in \mathbf{K}_m\}| + \epsilon\right)
        \end{split}  
        \label{eq:multiplicative}
      \end{equation}
    \normalsize

  \item PM$^+$(F): Additive Partial Match Model w/ Failing Tests
  \small
    \begin{equation}
      \begin{split}
        &\argmax_{i} \sum_{t \in \mathbf{T}_f} \fix{Pr}\left(M_{e_i} \mid F_{t}\right)\\
        &=\argmax_{i} \sum_{t \in \mathbf{T}_f} |\{m \in \mathbf{X}_{e_i} : t \in \mathbf{K}_m\}|
      \end{split}  
      \label{eq:additive}
    \end{equation}
    \normalsize
\end{itemize}

Intuitively, instead of counting exact matches, we want to aggregate scores
from the relationship between individual failing test cases and all mutants on 
a specific program element. PM$^*$(F) and PM$^+$(F) respectively aggregate individual
scores by multiplication and addition. Note that the PM$^*$(F) model requires a 
small positive quantity $\epsilon$ to prevent the value of the entire formula
from being zero when there exist one or more terms that evaluate to zero:
the value of $\epsilon$ does not affect the ranking.

Similarly to the case of EM models, we can also include the information of test cases that pass under the actual fault. These two models are called PM$^*$(F+P) and PM$^+$(F+P), and defined as follows:

\begin{itemize}
  \item PM$^*$(F+P): Multiplicative Partial Match Model w/ All Tests
  
  \small
  \begin{equation}
    \begin{split}
      &\argmax_{i} \left(\prod_{t \in \mathbf{T}_f} \left(\fix{Pr}\left(M_{e_i} \mid F_{t}\right) + \epsilon\right) \prod_{t \in \mathbf{T}_p} \left(\fix{Pr}\left(M_{e_i} \mid P_{t}\right) + \epsilon\right)\right)\\
      &=\argmax_{i} \left(\prod_{t \in \mathbf{T}_f} \left(|\{m \in \mathbf{X}_{e_i} : t \in \mathbf{K}_m\}| + \epsilon\right)\right.\\
      &\qquad\qquad\qquad\qquad\qquad\left.\times\prod_{t \notin \mathbf{T}_f} \left(|\{m \in \mathbf{X}_{e_i} : t \notin \mathbf{K}_m\}| + \epsilon\right)\right)\\
      &=\argmax_{i} \prod_{t \in \mathbf{T}} \left(|\{m \in \mathbf{X}_{e_i} : t \in \mathbf{T}_f \iff t \in \mathbf{K}_m \}| + \epsilon\right)
    \end{split}
    \label{eq:multiplicative_fp}
  \end{equation}
  \normalsize
  
  \item PM$^+$(F+P): Additive Partial Match Model w/ All Tests
  \small
  \begin{equation}
    \begin{split}
      &\argmax_{i} \left(\sum_{t \in \mathbf{T}_f} \left(\fix{Pr}\left(M_{e_i} \mid F_{t}\right)\right) + \sum_{t \in \mathbf{T}_p} \left(\fix{Pr}\left(M_{e_i} \mid P_{t}\right)\right)\right)\\
      &=\argmax_{i} \sum_{t \in \mathbf{T}} |\{m \in \mathbf{X}_{e_i} : t \in \mathbf{T}_f \iff t \in \mathbf{K}_m \}|
    \end{split}
    \label{eq:additive_fp}
  \end{equation}
  \normalsize
\end{itemize}

\subsection{Ranking Models Using Classifiers}
\label{sec:ranking_model_classifiers}

Scores from the Bayesian inference models described in Section~\ref{sec:exact_matching}
and~\ref{sec:partial_matching} are directly computed from the kill 
matrix, and requires virtually no additional analysis cost when scores 
are needed to be computed. However, all these models simply rely on counting 
matches between test results under the actual fault and kill matrix from the 
ahead-of-time mutation analysis. 

To investigate if more sophisticated statistical inference techniques can
improve the accuracy of \name, we apply both linear and non-linear classifiers
to build predictive models. These classifiers take the test results as input,
and yield the most suspicious method, as well as the suspiciousness score of
each method as output. Note that, in our study, we selected program element
$e$ to be the method where the mutant is located, as a method-level FL is one of the preferred granularity levels for FL by the developers~\cite{Kochhar:2016aa} and it eases the comparison with previous work whose results are only present in the method-level.


Let $\alpha_\mathbf{T_i}$ denote a 0-1 vector of the test results of
$\mathbf{T_i}$, where 0 indicates that test case fails, and 1 indicates that
test case passes. We first build a training set using the kill matrix
$\mathbf{K}$: test results per mutant $\mathbf{T_i}$ are transformed into
$\alpha_\mathbf{T_i}$, and the class is labelled based on the method where the
mutant is located.
We train representative linear and non-linear classifiers using Logistic
Regression (LR) and Multi-Layer Perceptron
(MLP)~\cite{yu2011dual,hinton1990connectionist}. For our study, we use a vanilla
MLP that consists of one input layer, one hidden layer with 50 neurons, and one
output layer. In the serving phase, we use the
suspiciousness score of each program element, which is obtained before the model
computes the most suspicious method. Only using the observed failures, we can
compose 0-1 vectors (i.e., LR(F) and MLP(F)), or compose 0-1 vectors by
including the information of passing tests (i.e., LR(F+P) and MLP(F+P)). Note
that, unlike the Bayesian inference models described in  
Section~\ref{sec:exact_matching} and~\ref{sec:partial_matching}, training these 
classifiers requires additional analysis cost to \name, although the training cost of 
these models is much lower than the cost of mutation analysis.


\subsubsection{Ranking Model Based on Probabilistic Coupling}
All \name models are intuitive based on the concept of coupling effect. We
now exploit the strict definition of the mutant-fault coupling to build a new
model based on probabilistic coupling~\cite{chen2020revisiting}, which aims to
quantify the degree of coupling between mutants and faults. A mutant is said to
be \emph{coupled} to a fault if the tests that kill the mutant can also kill
(i.e., detect) the fault. That is, if the mutant \emph{dynamically subsumes} the
fault, we call the mutant is perfectly coupled to the
fault~\cite{kurtz2014mutant}. Adopting this subsumption relationship for FL, we
focus on the mutants coupled with the faults: the stronger mutants are coupled
to a fault, the more likely they are to be closer to the fault.

Probabilistic coupling extends the notion of perfect coupling by
considering a probabilistic detection of faults in order to precisely estimate
the fault-revealing capability of mutants~\cite{chen2020revisiting}. Even if the
mutant does not subsume the fault, it can be \emph{probabilistically} coupled to
the faults if at least \emph{some} tests that kill the mutant are also able to
detect the faults. In the following, we illustrate the example cases of the
perfect coupling and probabilistic coupling, as well as how we model them as
scores to rank the likely fault locations.

\begin{enumerate}
  \item Perfect coupling: Let us assume that $t_2$ and $t_3$ in
  Table~\ref{tab:kill_matrix} are fault revealing tests (i.e., failing tests).
  From the view of the subsumption, $m_2$ is perfectly coupled to the fault
  because it is detected by $t_3$ but the fault is detected by both $t_2$ and
  $t_3$.

  \item Probabilistic coupling: On the other hand, mutant $m_1$ does not
    subsume the fault (i.e., not perfectly coupled). However, as it is detected
    by $t_2$, there is a possibility that $m_1$ is coupled to the fault if
    $m_1$ is detected by $t_2$ and also the fault if detected by $t_2$.
\end{enumerate}

Based on these two notions, we define a new model called PC that rewards mutants
coupled to the faults. The score of each mutant is computed as follows:
\begin{equation}
  pc(m)= 
  \begin{cases}
      1 ,      & \text{if } \mathbf{K}_m \subseteq \mathbf{T}_f \\
      \frac{|\mathbf{K}_m \cap \mathbf{T}_f|}{|\mathbf{K}_m|},   & \text{if } \mathbf{K}_m \nsubseteq \mathbf{T}_f \text{ and } |\mathbf{K}_m \cap \mathbf{T}_f| > 0 \\
      0,       & \text{otherwise.}
  \end{cases}
\end{equation}

The full score (i.e., 1) is given to the perfectly coupled mutant, the partial
score is given to the probabilistically coupled mutant, and no score is given to
the decoupled mutant (i.e., 0). We aggregate the mutants in the same method by
summing up their scores to compute a method-level FL score. Note that the
notion of probabilistic coupling cannot be defined if we only consider the
failing tests. Therefore we only present PC(F+P) model as follows:

\begin{equation}
  \begin{split}
    \text{PC(F+P)}= \argmax_{i} \sum_{m \in \mathbf{X}_{e_i}} pc(m)
  \end{split}        
\end{equation}


\section{Experimental Design}
\label{sec:exp_design}
This section describes the design of our empirical evaluation, including the 
way we use \dfj benchmark, the research questions, as well as other 
environmental factors.

\subsection{Protocol}
\label{sec:protocol}

One foundational assumption of \name is that existing test cases can be fault revealing
also for future changes. That is, for future faults to which \name will be applied,
test cases that would reveal them are available at the time of
the ahead-of-time mutation analysis. We believe this is a likely scenario mainly
in two contexts: regression faults, which are defined as failures of existing
test cases, and pre-commit testing, for which developers depend on existing test
cases for a sanity check. \name is designed to reduce the cost of MBFL for these
scenarios.\footnote{Although we do note that the more mature a software system is 
and the stronger and more complete its test suite is, the more likely it is that 
these conditions hold and thus that the proposed approach can be useful.}

However, this makes realistic experiments on real-world data challenging since a
majority of failure triggering changes are not likely to have been committed to
the main branch of the Version Control System (VCS): one of the purposes of
Continuous Integration is to prevent such commits~\cite{Liang2018fv}.
Consequently, fault benchmarks, such as \dfj, contain faults that have been
reported externally (e.g., from issue tracking systems), and provide fault
revealing test cases that have been added to the VCS with the patch
itself~\cite{Just:2014aa}. This presents a challenge for the realistic
evaluation of \name in the context it was designed for. To address this
issue, we introduce three experimental protocols, Faulty Commit Emulation (FCE),
Test Existence Emulation (TEE), and Kill Matrix Prediction (KMP).

\subsubsection{Faulty Commit Emulation (FCE)}
\label{sec:artificial_scenario}

This scenario emulates a faulty \emph{commit} that would 
trigger failures of existing test cases simply by reversing a fix patch in 
\dfj. We take the fixed version ($V_{fix}$) in \dfj as the reference version and performs the mutation analysis, 
including the test cases from the same version. Subsequently, we reverse the 
fix patch, execute the same test cases, and try to localise the fault using
the results with \name.

We argue that this is more realistic than injecting mutation faults 
artificially to evaluate \name. Since mutants are exactly what \name uses to 
build its models, \name may unfairly benefit if evaluated using mutants as
faults. Instead, we emulate faulty commits using faults that some developers 
actually had introduced in real-world software. Existing work on test data 
generation has also used the fixed version as the reference version, against
which a test generation tool is applied. The reversed fix patch is then
used to emulate regression faults for the evaluation of the generated 
tests~\cite{Shamshiri2015ase,Just2014fse}. Our approach with FCE
is similar in the sense that we analyse the fixed version first,
then use the outcome to localise the emulated regression fault.


\subsubsection{Test Existence Emulation (TEE)}
\label{sec:realworld_scenario}

This scenario uses original faulty commits that led to the faulty versions 
($V_{bug}$) in \dfj, but simply \emph{pretends} that the \emph{fault revealing test cases
existed earlier}. We have checked whether the fault revealing test cases in \dfj
can be executed against versions that precede the actual faulty version.
Since system specifications evolve over time, executing a future test case
against past versions is not always successful: we have identified 28 previous
versions for which the future fault revealing test cases can be executed and 
\emph{do not fail}. We use these 28 versions as references, and use their
mutation analysis results to localise the corresponding faults that happened later.
Compared to FCE, 
TEE follows the ground truth code changes, and only assumes the earlier 
existence of fault revealing test cases. We use TEE to complement the FCE 
scenario. Specifically, TEE can evaluate whether training \name models with 
kill matrices of earlier versions degrades its 
localisation accuracy.

\subsubsection{Kill Matrix Prediction (KMP)}
\label{sec:simfl_scenario}

Lastly, we present a new scenario that is free from the pre-existence of
failing tests. The basic idea is to utilise a predicted kill matrix inferred by
\se~\cite{Kim2021ax}, a predictive mutation analysis technique producing an
entire kill matrix. Having learnt from the kill matrix computed from a past
version, \se can predict the kill matrix of the current version based on the
syntactic and semantic features in source code, test code, and mutant. Most
importantly, \se can predict the kill results of test cases that are introduced
after the reference version from whose kill matrix \se learnt from. That is,
this scenario using \se does not care about the pre-existence of failing tests.
For example, to locate the faults of Lang programs, we first take the oldest
version of them as the reference version, i.e., $V_{fix}$ of \texttt{Lang
65}\footnote{We denote version numbers of \texttt{Lang} programs by the
identifier numbers used in \dfj.}. Then using \se, we predict the kill matrix of
the faulty version which is more recent version than \texttt{Lang 65}, e.g.,
$V_{bug}$ of \texttt{Lang 30}. Next, we train the \name models based on the
predicted kill matrix of $V_{bug}$ of \texttt{Lang 30}, and produce a ranked
list of likely faulty locations. Note that we do use $V_{fix}$ of \texttt{Lang
65} to predict the faults of $V_{bug}$ of \texttt{Lang 65}, which is what FCE
does.

\subsubsection{Building Kill Matricies for FCE and TEE}
\label{sec:km_for_fce}
Building a full kill matrix requires huge computational cost: mutation 
analysis on all versions of Closure using Major exceeded our 24 hours 
timeout, and other subject programs also required significant amounts of
analysis time. To address this practical concern, for empirical evaluation,
we have constructed the kill matrix using only the 
\emph{relevant test cases} as defined by \dfj\footnote{See https://github.com/rjust/defects4j/tree/v1.3.1\#export-version-specific-properties}, 
which include the failing test cases as well as any passing test cases that 
makes the JVM to load at least one of the classes modified by the fault 
introducing commit.

Note that this procedure has been adopted strictly to reduce experimental cost.
Since we only have the kill matrix for the relevant test cases, models that 
use F+P test cases actually use the full set of relevant test cases. However, 
if construction of the full kill matrix is feasible, the same input used by 
\name in this paper is naturally available. The F+P models can be trained  
either using the full set of test cases (increased training cost but also
richer input information), or using the relevant test cases (relevancy
information is still cheaper than full coverage instrumentation).
We argue that, in general, the limitation to only the relevant test cases is a
conservative one and should reduce rather than improve the fault localisation
accuracy of \name since other test cases could also be informative for its
statistical models.

\subsubsection{Building Kill Matricies for KMP}
\label{sec:km_for_seshat}
Inputs for training \se include the full kill matrix and the tokens of
  source and test code, meaning that \se has the same problem raised in
  Section~\ref{sec:km_for_fce} that requires substantial costs for computing
  full kill matrix. Thus we exclude all versions of Closure that did not halt
  within a given timeout. Also, we exclude all versions of Time to obtain the full
  kill matrix using \emph{all tests}. Compared to FCE that uses \emph{relevant
  test cases} by \dfj, we believe that it is important for \se to be trained on
  a diverse and large set of tests to retain its predictive power on unseen source
  code and tests.

 We select the reference model as the oldest fixed version of each project,
  e.g., $V_{fix}$ of \texttt{Lang 65} or \texttt{Math 106}. They are used to
  train \se and later used to predict the full kill matrix of target buggy
  version. Please refer to the original paper of \se~\cite{Kim2021ax} for the
  prediction accuracy and required time of \se itself.

\subsubsection{Using Test Runtime Information}
\label{sec:runtime_info}

The use case of \name assumes that, while the actual mutation analysis can be
performed in advance, the inference models are trained after the observation 
of a failure (see \cref{fig:scenario}). In practice, the observation of the 
behaviour of the failing test cases can provide information that is beyond
the mutation analysis. Consequently, we exploit this additional information 
by collecting coverage reports of failing test cases using \texttt{Cobertura}.
We then exclude any methods and mutants that are not covered by the failing
test cases from model training and the final ranking.

\begin{table}[ht]
  \centering
  \caption{Subject Programs in \dfj \label{tab:subject}}  
  \scalebox{0.62}{
  \begin{tabular}{l|rrrrr}
  \toprule
  Subject & \# Faults & kLoC & \# Methods & \# Mutants & \# Test cases\\
  \midrule
  Commons-lang (Lang) & 65  & 50  & 1,527  & 21,178  & 2,245  \\
  JFreeChart (Chart)         & 26  & 132  & 4,903  & 75,985  & 2,205  \\
  Joda-Time (Time)           & 27  & 105  & 1,946  & 21,689  & 4,130  \\
  Closure compiler (Closure) & 133 & 216  & 5,038  & 58,515  & 7,927  \\
  Commons-math (Math) & 106 & 104  & 2,713  & 79,428  & 3,602  \\
  \midrule
  Total                      & 357 & 607 & 16,126 & 256,792 & 20,109 \\
  \bottomrule
  \end{tabular}  
  }
\end{table}

\subsection{Subject Programs}
\label{sec:subject}

In our study, we use 357 versions of five different programs from the \dfj
version 1.3.1. They provide reproducible and isolated faults of real-world
programs. Table~\ref{tab:subject} summarises the subject programs we used with
the average number of generated mutants, methods, lines of code, and test cases
across all faults belonging to each subject respectively. We could not include
Mockito as we failed to compile the majority of its versions and their mutants
using the build script provided by \dfj on Docker containers. Moreover, we
include the faults newly introduced in \dfj version 2.0.0, which consists of 401
new faults. In Section~\ref{sec:RQ1_dfj2}, we report the results of using these
new faults.




\subsection{Research Questions}
\label{sec:rqs}

\vspace{0.5em}

\noindent\textbf{RQ1. Localisation Effectiveness on FCE:} \textit{Does the
	      models of \name produce accurate fault localisation compared to the
	      state-of-the-art FL techniques?} RQ1 is answered by computing the
	      standard evaluation metrics on the eight models of \name under the FCE
	      scenario outlined in~Section~\ref{sec:artificial_scenario}. We compare
	      \name with two MBFL techniques (MUSE and Metallaxis), two SBFL
	      techniques (Ochiai and DStar), and two learning-to-rank based FL
	      techniques (TraPT and FLUCCS). Moreover, we further evaluate ...
        
\vspace{0.5em}

\noindent\textbf{RQ2. Model Viability on TEE:} \textit{How well does \name hold up when
	      applied using models built earlier?} RQ2 is answered by computing the
	      standard evaluation metrics using prior models built under the TEE
	      scenario outlined in~Section~\ref{sec:realworld_scenario}.
        
\vspace{0.5em}

\noindent\textbf{RQ3. Localisation Effectiveness on KMP:} \textit{How well
does \name localise faults when it uses the predicted kill matrix inferred by
\se?} To answer RQ3, we conduct the same experiment with RQ1, but with the kill
matrices  predicted by \se. We report standard evaluation metrics and investigate
the effects of using predicted kill matrices.

\vspace{0.5em}

\noindent\textbf{RQ4. Efficiency:} \textit{How efficient is
	      \name compared to other MBFL techniques?} Despite being an MBFL
	      technique, \name can amortise the time needed to localise faults. With this RQ, we
        directly compare the execution time of \name to MUSE and Metallaxis, 
        as well as the time needed to run a whole test suite once.

\vspace{0.5em}

\noindent\textbf{RQ5. Effects of Sampling:} \textit{What is the impact of
	      mutation sampling to the effectiveness of \name?} Since the cost of
	      mutation analysis is the major component of the cost of \name, we
	      investigate how much impact different mutation sampling rates have. We
	      evaluate two different sampling techniques: uniform random sampling,
	      which samples from the pool of all mutants uniformly, and stratified
	      sampling, which samples numbers of mutants as uniformly as possible
	      across methods.

\vspace{0.5em}

\noindent\textbf{RQ6. Effects of Subsumed Mutants:} \textit{What is the
impact of subsumed mutants to \name models?} As we suspect that the subsumed
mutants would inflate the method-level aggregation of \name scores, we study the
impact of removing the subsumed mutants by comparing $acc@n$ scores with RQ1
results.

\subsection{Evaluation Metrics and Tie Breaking}
\label{sec:metrics}
We use three standard evaluation metrics:
\begin{itemize}
  \item $acc@n$: counts the number of faults located within top $n$ ranks.
                We report $acc@1$, $acc@3$, $acc@5$, and $acc@10$. If a fault is patched across multiple methods, we take the highest ranked method to compute $acc@n$.

  \item \noindent$wef$: approximates the amount of efforts wasted by developers
  while investigating non-faulty methods that are ranked higher than the faulty method.

\item \noindent Mean Average Precision (MAP): measures the mean of the average
precision values for a group of all faults. For each fault, when each faulty
program elements are ranked at $R = \{r_1, \dots, r_n\}$, where $r_i$ is the
higher rank than $r_{i+1}$, the average precision is calculated as
$\frac{1}{|R|}\sum_{i=1}^{n} \frac{i}{r_i}$. The faulty methods not retrieved
get a precision score of zero.

\end{itemize}

If multiple program elements have the same score, resulting in the same rank, we break the tie using max tie breaker that places all program elements with the same score at the lowest rank.


\begin{table*}[ht]
  \centering
  \caption{Effectiveness of \name models on FCE scenario.\label{tab:RQ1_total_ranks}}
  \scalebox{0.77}{
    \begin{tabular}{l|lr|rrrr|r|r||l|lr|rrrr|r|r}
      \toprule
      Model & Project  & Total & \multicolumn{4}{c|}{$acc$} & $wef$ & MAP & Model & Project  & Total & \multicolumn{4}{c|}{$acc$} & $wef$ & MAP\\
      &   & Studied & @1 & @3 & @5 & @10 & med  & & &  & Studied & @1 & @3 & @5 & @10 & med & \\
      \midrule
      \multirow{6}{*}{\shortstack[l]{EM\\(F)}} & Lang & 65 & 35 & 45 & 47 & 48 & \textbf{0.0} &  0.6176 & \multirow{6}{*}{\shortstack[l]{EM\\(F+P)}} & Lang & 65 & 36 & 41 & 43 & 44 & \textbf{0.0} & 0.5922 \\
      & Chart & 26 & \textbf{6} & \textbf{11} & \textbf{13} & 15 & 5.0 &   0.3294 & & Chart & 26 & 6 & 9 & 10 & 11 & 27.0 & 0.2917 \\
      & Time & 27 & \textbf{4} & 9 & 9 & 13 & 8.5 &  0.2451 & & Time & 27 & 10 & 13 & 14 & 15 & 3.0 & 0.3819 \\
      & Closure & 133 & 10 & 31 & 41 & 57 & 17.0 &   0.1753 & & Closure & - & - & - & - & - & - & - \\
      & Math & 106 & 22 & 43 & 53 & 71 & 4.0 & 0.3404 & & Math & 106 & 32 & 45 & 47 & 49 & 3.0  & 0.4098 \\ \cmidrule{2-9} \cmidrule{11-18} 
      & Total & 357 & 77 & 139 & 163 & 204 & & & & Total & 224 & 84 & 108 & 114 & 119 & & \\
      \midrule
      \multirow{6}{*}{\shortstack[l]{PM$^*$\\(F)}} & Lang & 65 & 38 & 47 & 51 & 53 & \textbf{0.0} & 0.6732 & \multirow{6}{*}{\shortstack[l]{PM$^*$\\(F+P)}} & Lang & 65 & 27 & 36 & 37 & 42 & 1.0  & 0.5264 \\
      & Chart & 26 & \textbf{6} & \textbf{11} & \textbf{13} & 16 & 5.0 & 0.3562 & & Chart & 26 & 7 & 9 & 12 & 14 & 6.0  & 0.3598 \\
      & Time & 27 & \textbf{4} & \textbf{10} & 10 & 13 & 8.0  & 0.2549 & & Time & 27 & 1 & 3 & 4 & 12 & 16.0  & 0.1172 \\
      & Closure & 133 & 11 & 36 & 50 & 66 & 9.5  & 0.1982 & & Closure & - & - & - & - & - & - & -\\
      & Math & 106 & 23 & \textbf{47} & 59 & 77 & 3.5 & 0.3753 & & Math & 106 & 14 & 26 & 33 & 42 & 12.0 & 0.2460 \\ \cmidrule{2-9} \cmidrule{11-18} 
      & Total & 357 & 82 & 151 & 183 & 225 & & & & Total & 224 & 49 & 74 & 86 & 110 & & \\
      \midrule
      \multirow{6}{*}{\shortstack[l]{PM$^+$\\(F)}} & Lang & 65 & 40 & 48 & 52 & 53 & \textbf{0.0} & 0.6977 & \multirow{6}{*}{\shortstack[l]{PM$^+$\\(F+P)}} & Lang & 65 & 19 & 31 & 33 & 37 & 2.0 & 0.4291 \\
      & Chart & 26 & \textbf{6} & 10 & \textbf{13} & \textbf{19} & \textbf{4.0}  & \textbf{0.3697} & & Chart & 26 & 5 & 9 & 12 & 13 & 8.0  & 0.2712 \\
      & Time & 27 & \textbf{4} & \textbf{10} & 10 & 13 & 8.0  & 0.2564 & & Time & 27 & 0 & 2 & 3 & 5 & 40.5  & 0.0616 \\
      & Closure & 133 & \textbf{12} & \textbf{41} & \textbf{52} & 65 & 11.0  & 0.2005 & & Closure & - & - & - & - & - & - & -\\
      & Math & 106 & 24 & 46 & 59 & 77 & 4.0  & 0.3845 & & Math & 106 & 9 & 15 & 20 & 29 & 23.0  & 0.1574 \\ \cmidrule{2-9} \cmidrule{11-18} 
      & Total & 357 & 86 & \textbf{155} & \textbf{186} & \textbf{227} & & & & Total & 224 & 33 & 57 & 68 & 84 & & \\  
      \midrule
      \multirow{6}{*}{\shortstack[l]{LR\\(F)}} & Lang & 65 & \textbf{41} & 49 & \textbf{53} & \textbf{55} & \textbf{0.0}  & \textbf{0.7179} & \multirow{6}{*}{\shortstack[l]{LR\\(F+P)}} & Lang & 65 & 40 & 49 & 51 & 53 & \textbf{0.0} & 0.7017 \\
      & Chart & 26 & 5 & 9 & 12 & 14 & 6.0 & 0.3175 & & Chart & 26 & 8 & 14 & 14 & 16 & \textbf{2.0}  & 0.4194 \\
      & Time & 27 & \textbf{4} & \textbf{10} & \textbf{12} & \textbf{14} & 5.5  & 0.2668 & & Time & 27 & 8 & 14 & 17 & 19 & 2.0 & 0.4094 \\
      & Closure & 133 & \textbf{12} & 37 & 50 & \textbf{68} & \textbf{9.0}  & \textbf{0.2074} & & Closure & - & - & - & - & - & - & -\\
      & Math & 106 & \textbf{28} & \textbf{47} & 59 & 75 & \textbf{3.0} & \textbf{0.3976} & & Math & 106 & 32 & 43 & 47 & 51 & 3.0 & 0.4066 \\ \cmidrule{2-9} \cmidrule{11-18} 
      & Total & 357 & \textbf{90} & 152 & \textbf{186} & 226 & & & & Total & 224 & 88 & 120 & 129 & 139 & &  \\
      \midrule
      \multirow{6}{*}{\shortstack[l]{MLP\\(F)}} & Lang & 65 & 39 & \textbf{51} & \textbf{53} & \textbf{55} & \textbf{0.0} & 0.7052 & \multirow{6}{*}{\shortstack[l]{MLP\\(F+P)}} & Lang & 65 & \textbf{48} & \textbf{55} & \textbf{56} & \textbf{56} & \textbf{0.0}  & \textbf{0.7882} \\
      & Chart & 26 & 5 & 10 & 12 & 15 & 6.0 & 0.3319 & & Chart & 26 & 9 & 13 & 15 & 19 & 2.0  & 0.4477 \\
      & Time & 27 & \textbf{4} & \textbf{10} & \textbf{12} & \textbf{14} & \textbf{5.0} & \textbf{0.2710} & & Time & 27 & \textbf{11} & \textbf{16} & \textbf{18} & \textbf{24} & \textbf{1.0} & \textbf{0.4847} \\
      & Closure & 133 & 11 & 33 & 41 & 60 & 12.0  & 0.1888 & & Closure & - & - & - & - & - & - & - \\
      & Math & 106 & 26 & 46 & \textbf{62} & \textbf{79} & \textbf{3.0} & 0.3941 & & Math & 106 & \textbf{45} & \textbf{61} & 70 & \textbf{82} & \textbf{1.0}  & \textbf{0.5194} \\ \cmidrule{2-9} \cmidrule{11-18} 
      & Total & 357 & 85 & 150 & 180 & 223 & & & & Total & 224 & \textbf{113} & \textbf{145} & \textbf{159} & \textbf{181} & & \\
      \midrule      
      
      \multicolumn{9}{c||}{} & \multirow{6}{*}{\shortstack[l]{PC\\(F+P)}} & Lang & 65 & 42 & 51 & 51 & 54 & \textbf{0.0} & 0.7323 \\
      \multicolumn{9}{c||}{} &  & Chart & 26 & \textbf{10} & \textbf{16} & \textbf{17} & \textbf{22} & \textbf{1.0} & \textbf{0.5389} \\
      \multicolumn{9}{c||}{} &  & Time & 27 & 7 & 15 & \textbf{18} & 21 & 2.0 & 0.4133 \\
      \multicolumn{9}{c||}{} &  & Closure & - & - & - & - & - & - & - \\
      \multicolumn{9}{c||}{} &  & Math & 106 & 37 & \textbf{61} & \textbf{71} & 81 & 1.5 & 0.4934 \\  \cmidrule{11-18}
      \multicolumn{9}{c||}{} &  & Total & 224 & 96 & 143  & 157  & 178 &  &  \\
      \bottomrule
      \end{tabular}
  }
\end{table*} 

\subsection{Mutation Tool and Operators}
\label{sec:mutation_tool_and_operator}

In the study, we use Major version 1.3.4~\cite{Just2011gq} as our mutation analysis tool,
and choose all mutation operators in Major.
Note that some operators had to be turned off for specific classes so
that Major does not generate an exceptionally large number of mutants.\footnote{Due to the internal design of Major, some classes that
yield too many mutants may lead to the violation of bytecode length limit 
imposed by Java compiler. See \url{https://github.com/rjust/defects4j/issues/62} 
for technical details.}



\section{Results}
\label{sec:result}


\subsection{Localisation Effectiveness on FCE (RQ1)}
\label{sec:result_RQ1}

We start by comparing different \name models. Subsequently, using the best 
\name model, we compare \name to the state-of-the-art fault localisation 
techniques.

\subsubsection{Comparison Between \name Models}

Table~\ref{tab:RQ1_total_ranks} shows the results of each evaluation metric 
for all studied faults, following the FCE scenario. The numbers $X(Y)$ in the 
column ``Total Studied'' represent the number of faults that we can localise 
$(X)$, and the number of faults provided by \dfj $(Y)$. Evaluation metric values 
representing the
best outcome (i.e., the largest $acc@n$ and MAP, and the smallest $wef$) are typeset
in bold. See 
Section~\ref{sec:subject} for the details of exclusion criteria we used: note 
that more faults are excluded from the study of F+P models shown on the right.

Overall, MLP(F+P) shows the best performance in terms of $acc@n$ metrics,
placing 48 out of 61 faults at the first place for Lang, and 45 out of 91 
faults at the first place for Math. Considering that MLP(F+P) is evaluated
on fewer faults (203) than MLP(F) (348), the result suggests that MLP(F+P)
shows better performance on average.

We argue that including results of 
passing tests gives richer information when compared to only using results of 
failing tests. However, we also note that only MLP significantly benefits from
the additional information: MLP(F+P) places 28 more faults at the first rank than 
MLP(F). Two linear models, LR(F) and LR(F+P), on the other hand, do not show 
any significant difference in performance. This suggests that exploiting this 
information requires more sophisticated, non-linear inference methods. 

The reason that PM$^+$(F) shows comparable results to MLP(F) may be that
it is relatively easy to simply count the matching patterns of failing tests, 
which are much rarer than passing tests. We also note that PM$^*$(F) and 
PM$^+$(F) both produce better results than EM(F), suggesting that partial
matches are better than exact matches. This is because even the fault revealing
test case may not be able to kill all mutations applied to the location of
the fault. In such a case, 
the EM(F) model will lose the information, while the PM(F) models will benefit from
other killed mutants from the same location.

Finally, the addition of passing test information to PM models actually
degrades the performance significantly, as the metrics for PM$^*$(F+P) and 
PM$^+$(F+P) show. Partially matching test cases that did not fail against the 
faulty version with test cases that did not kill mutants at the location of 
the fault will directly dilute the signal, as failing tests and killed mutants 
are likely to provide more information about the location of the fault in 
general.


\begin{table}[!ht]
  \centering  
  \caption{Evaluation of \name models on the faults introduced in \dfj 2.0.0.}
  \label{tab:d4j20}  
  \scalebox{0.8}{
    \begin{tabular}{l|l|rrrr}
      \toprule
      Model & Total   & $acc@1$ & $acc@3$ & $acc@5$ & $acc@10$ \\
      \midrule
      EM(F) & 295 & 42 & 80 & 101 & 126 \\
      PM$^*$(F) & 295 & 49 & 85 & 107 & 133 \\
      PM$^+$(F) & 295 & 45 & 84 & 105 & 130 \\
      LR(F) & 295 & 48 & 93 & 110 & 145 \\
      MLP(F) & 295 & 51 & 87 & 109 & 139 \\
      \midrule
      EM(F+P) & 295 & 71 & 109 & 139 & 158 \\
      PM$^*$(F+P) & 295 & 43 & 84 & 103 & 137 \\
      PM$^+$(F+P) & 295 & 39 & 73 & 100 & 130 \\
      LR(F+P) & 295 & 68 & 117 & 136 & 173 \\
      MLP(F+P) & 295 & \textbf{85} & \textbf{139} & \textbf{168} & \textbf{194} \\
      PC(F+P) & 295 & 78 & 127 & 161 & 188 \\
    \bottomrule
      \end{tabular}
    }
\end{table}

\subsubsection{Evaluation on \dfj 2.0.0}
\label{sec:RQ1_dfj2}
We expand our assessment of \name to include new 401 faults that have been
added to \dfj 2.0.0, which span 11 new Java projects. Among these faults, Major
mutation tool could successfully run on 295 faults, which are the ones we use
for evaluation.

The results with $acc@n$ metrics are shown in Table~\ref{tab:d4j20}.
  Consistent with our previous findings, MLP(F+P) model continues to outperform
  the other models across all $acc@n$ metrics. Furthermore, the results reaffirm the
  effects of incorporating passing tests, as demonstrated by the improved
  performance of MLP(F+P) and LR(F+P) models, and degraded performance of
  PM$^*$(F+P) and PM$^+$(F+P) over their F-only counterparts. All in all, this extended
  evaluation demonstrates the validity of our study on \dfj.

\begin{figure}[ht]
    \centering    
    \includegraphics[width=.5\textwidth]{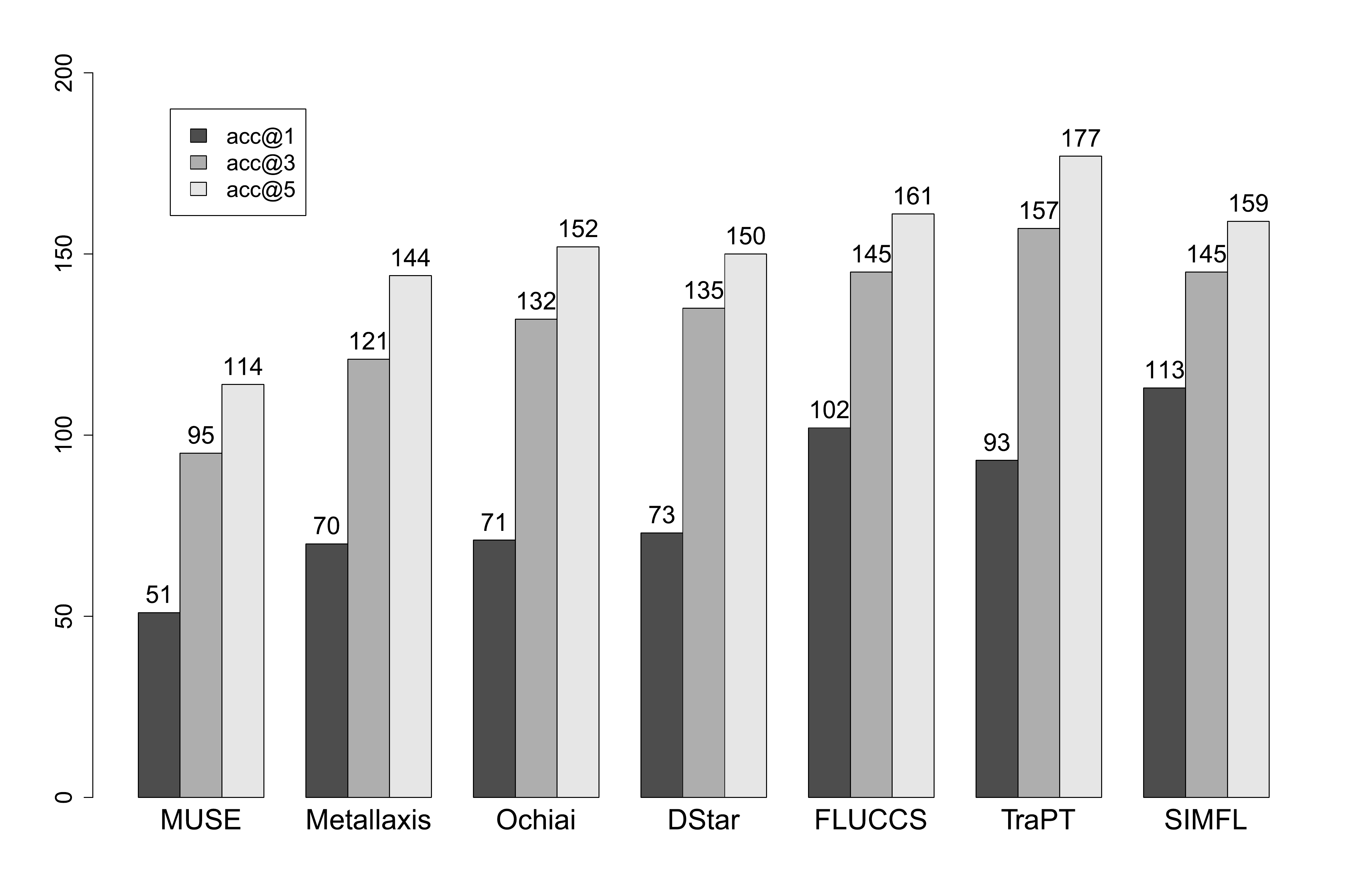}
    \caption{Comparison to other FL techniques:
    $acc@n$ metric values without counts of Closure\label{fig:RQX_rank_comparison}}
\end{figure}

\subsubsection{Comparison to Other FL Techniques}
\label{sec:comparison}

To gain some insights into the trade-off between amortised modelling efforts 
and localisation accuracy, we compare the method-level
fault localisation results of the state-of-the-art MBFL and SBFL
techniques, the result of which is shown in \cref{fig:RQX_rank_comparison}. 
We obtained the performance of the each model (i.e., $acc@n$) on \dfj
from the literatures, and artefact of Zou et al.~\cite{zou2019empirical}.
Based on the results of the comparison between \name models, we choose
MLP(F+P) to represent \name. However, since Closure has been excluded from 
the evaluation of F+P models, we have excluded Closure from the results of other techniques for a fair comparison.

\cref{fig:RQX_rank_comparison} shows that MLP(F+P) is better than other 
techniques in terms of $acc@1$, but TraPT performs better in terms of $acc@3$ and $acc@5$.
Although \name does not make use of learning-to-rank technique to boost performance
by fully including runtime information or suspiciousness scores of other FL techniques,
\name localises faults at the first rank better than others, and shows
comparable results to the learning-to-rank techniques: FLUCCS and TraPT.

\begin{tcolorbox}[boxrule=0pt,frame hidden,sharp corners,enhanced,borderline north={1pt}{0pt}{black},borderline south={1pt}{0pt}{black},boxsep=2pt,left=2pt,right=2pt,top=2.5pt,bottom=2pt]
\textbf{Answer to RQ1}: \name can localise faults 
accurately compared to the existing techniques: \name places up to 25.86\%
(90 of 348 for LR(F)) of studied faults at the first rank using F models,
and 55.67\% (113 of 203 for MLP(F+P)) of studied faults at the first rank using F+P models.
\end{tcolorbox}


\subsection{Model Viability on TEE (RQ2)}
\label{sec:result_RQ2}

\begin{table*}[ht]
  \centering
  \caption{Viability of F models using TEE scenario. The ranks that do not have the same ranks with FCE are typeset in bold. \label{tab:RQ2}}
  \scalebox{0.9}{
  \begin{tabular}{ll|lp{0.6cm}p{0.6cm}p{0.6cm}p{0.6cm}p{0.7cm}|ll|lp{0.6cm}p{0.6cm}p{0.6cm}p{0.6cm}p{0.7cm}}
  \toprule
  \multicolumn{2}{l|}{Fault}    & Commit             & \multicolumn{5}{c|}{Rank}                                                            & \multicolumn{2}{l|}{Fault} & Commit             & \multicolumn{5}{c}{Rank} \\
  & & ($\Delta$rev.) & EM                    & PM$^*$   & PM$^+$   & LR                    &  MLP        & & & ($\Delta$rev.) & EM       & PM$^*$   & PM$^+$   & LR       & MLP      \\ \midrule
  \parbox[t]{2mm}{\multirow{18}{*}{\rotatebox[origin=c]{90}{Closure}}} & 21  & FCE Rank           & 2                        & 2           & 2           & 2                        &  2           & \parbox[t]{2mm}{\multirow{18}{*}{\rotatebox[origin=c]{90}{Math}}} & 46  & FCE Rank           & 188         & 188         & 188         & 47          & 85          \\
          &     & 32a12ba (2)        & 2                        & 2           & 2           & 2                        &  2           &      &     & bbb5e1e (1)        & 188         & 188         & 188         & \textbf{35} & \textbf{47} \\
          &     & 43a5523 (3)        & 2                        & 2           & 2           & 2                        &  2           &      &     & 37680e2 (2)        & 188         & 188         & 188         & \textbf{35} & \textbf{47} \\ \cmidrule{2-8}
          & 61  & FCE Rank           & 7                        & 4           & 5           & 5                        &  6           &      &     & 1861674 (3)        & 188         & 188         & 188         & \textbf{35} & \textbf{27} \\ 
          &     & f5529dd (3)        & 7                        & 4           & 5           & 5                        &  6           &      &     & f0b12de (4)        & 188         & \textbf{1}  & \textbf{1}  & \textbf{3}  & \textbf{1}  \\
          &     & b12d1d6 (4)        & 7                        & 4           & 5           & 5                        &  6           &      &     & 8581b76 (5)        & 188         & 188         & 188         & \textbf{35} & \textbf{41} \\ \cmidrule{10-16}
          &     & 245362a (7)        & 7                        & 4           & 5           & 5                        &  \textbf{7}  &  & 89  & FCE Rank           & 13          & 13          & 13          & 8           & 7           \\
          &     & 8abd1d9 (8)        & 7                        & 4           & 5           & 5                        &  \textbf{8}  &      &     & 43336b0 (1)        & \textbf{12} & \textbf{12} & \textbf{12} & \textbf{2}  & \textbf{3}  \\
          &     & 37b0e1b (9)        & 7                        & 4           & 5           & 5                        &  6           &      &     & cdd62a0 (2)        & \textbf{14} & \textbf{14} & \textbf{14} & \textbf{2}  & \textbf{5}  \\ \cmidrule{2-8}
          & 62  & FCE Rank           & 1                        & 1           & 1           & 1                        &  1           &      &     & 90439e5 (3)        & 13          & 13          & 13          & 8           & \textbf{11} \\
          &     & 245362a (2)        & 1                        & 1           & 1           & 1                        &  1           &      &     & 36a8485 (4)        & 13          & 13          & 13          & 8           & \textbf{13} \\
          &     & 8abd1d9 (3)        & 1                        & 1           & 1           & 1                        &  1           &      &     & dbe7842 (5)        & 13          & 13          & 13          & 8           & 7           \\
          &     & 37b0e1b (4)        & 1                        & 1           & 1           & 1                        &  1           &      &     & d84a587 (6)        & 13          & 13          & 13          & 8           & \textbf{12} \\\cmidrule{2-8}
          & 115 & FCE Rank           & 14                       & 22          & 24          & 19                       &  11          &      &     & d27e072 (7)        & 13          & 13          & 13          & 8           & \textbf{10} \\
          &     & b9262dc (5)        & 14                       & \textbf{19} & \textbf{22} & \textbf{18}              &  \textbf{13} &      &     & 3590bdc (8)        & 13          & 13          & 13          & 8           & \textbf{8}  \\
          &     & 911b2d6 (6)        & 14                       & 22          & 24          & 19                       &  \textbf{12} &      &     & 6b108c0 (9)        & 13          & 13          & 13          & 8           & \textbf{13} \\\cmidrule{2-8}
          & 120 & FCE Rank           & 7                        & 7           & 7           & 6                        &  6           &      &     & 9c55428 (10)       & 13          & 13          & 13          & 8           & \textbf{12} \\
          &     & 2aee36e (3)        & \textbf{24}              & \textbf{24} & \textbf{24} & \textbf{15}              &  \textbf{16} &      &     &                    &             &             &             &             &             \\

  \bottomrule
  \end{tabular}}
\end{table*}

Following the TEE scenario described in Section~\ref{sec:realworld_scenario},
we seek reference versions preceding the faulty version, i.e., the versions
before the faulty version that pass all test cases
of the fixed program, including the fault revealing test cases. Assuming that
more recent versions are more likely to serve as references, given a faulty version
$n$, we check $n-1$, \dots, $n-10$, $n-20$, and $n-30$ previous program versions,
as it is impractical to inspect all of them. Starting from 357 faulty versions of
subject programs, we found 28 preceding reference versions 
that correspond to seven different faulty versions. We have trained 
five F models on each of the 28 reference versions to localise the fault in the faulty version, resulting in 140 rankings based on TEE scenario.
Note that we did not consider F+P models on these reference versions because they require
more than 24 hours for mutation analysis, as described in Section~\ref{sec:subject}.

Table~\ref{tab:RQ2} shows the rank of the faulty method for each F model
built on each preceding reference version. Out of 140 TEE based rankings produced by F models, 103 are identical to the corresponding FCE ranking. One notable exception
is Math 46 (f0b12de) that shows a significant improvement over the FCE scenario rank. 
We have manually examined the kill matrix of this reference version, and found
that some mutants in the future faulty method have been additionally killed 
due to timeout (enforced by Major itself), contributing to the high rank
(these mutants were not killed
in other preceding reference versions of Math 46). We suspect that this is due 
to the non-determinism in the process of building the kill matrix: the mutation
may have brought in flakiness that has been removed for the original program.
We study the impact of different kill reasons in Section~\ref{sec:kill_reason_filtering},
and furthermore discuss this as one of the threats to internal validity in \cref{sec:threats}.

Note that, in all cases, the $\Delta$ revision values are ten or fewer,
meaning that mutation analysis results from within ten past commits have been
used to perform the localisation. Since the failing test cases in \dfj are
typically added against the buggy version, it is often impossible to execute it
against a much older version of the same project, requiring a fairly recent kill
matrix. We expect that, if the failing test case is an existing regression test
that has remained executable for a long period of time, \name will be more
stable for longer durations. Our KMP scenario, which we will discuss in the next
section, is also developed to cater for cases in which the mutation analysis
results cannot be easily obtained for the failing test cases.

\begin{tcolorbox}[boxrule=0pt,frame hidden,sharp corners,enhanced,borderline north={1pt}{0pt}{black},borderline south={1pt}{0pt}{black},boxsep=2pt,left=2pt,right=2pt,top=2.5pt,bottom=2pt]
\textbf{Answer to RQ2}: The performances of \name using models built with preceding 
reference versions tend to be stable when compared to the FCE results: only 19
out of 140 cases show degraded performance since we
used less recent mutation analysis results.
\end{tcolorbox}

\begin{table*}[!ht]
  \centering  
  \caption{Effectiveness of \name models on KMP scenario, using all tests.
  \label{tab:seshat_minimized}}  
  \scalebox{1.0}{
    \begin{tabular}{l|l|rr|rr|rr|rr}
      \toprule
      Model & Total   & \multicolumn{2}{c|}{$acc@1$} & \multicolumn{2}{c|}{$acc@3$} & \multicolumn{2}{c|}{$acc@5$} & \multicolumn{2}{c}{$acc@10$} \\
      \midrule
      EM(F) & 194 & 57 & {\footnotesize \textcolor{minus}{$-20$}} & 95 & {\footnotesize \textcolor{minus}{$-44$}} & 108 & {\footnotesize \textcolor{minus}{$-55$}} & 122 & {\footnotesize \textcolor{minus}{$-82$}} \\
PM$^*$(F) & 194 & 61 & {\footnotesize \textcolor{minus}{$-21$}} & 101 & {\footnotesize \textcolor{minus}{$-50$}} & 114 & {\footnotesize \textcolor{minus}{$-69$}} & 131 & {\footnotesize \textcolor{minus}{$-94$}} \\
PM$^+$(F) & 194 & 60 & {\footnotesize \textcolor{minus}{$-26$}} & 104 & {\footnotesize \textcolor{minus}{$-51$}} & 119 & {\footnotesize \textcolor{minus}{$-67$}} & 134 & {\footnotesize \textcolor{minus}{$-93$}} \\
LR(F) & 194 & 63 & {\footnotesize \textcolor{minus}{$-27$}} & 104 & {\footnotesize \textcolor{minus}{$-48$}} & 116 & {\footnotesize \textcolor{minus}{$-70$}} & 134 & {\footnotesize \textcolor{minus}{$-92$}} \\
MLP(F) & 194 & 66 & {\footnotesize \textcolor{minus}{$-19$}} & 106 & {\footnotesize \textcolor{minus}{$-44$}} & 120 & {\footnotesize \textcolor{minus}{$-60$}} & 140 & {\footnotesize \textcolor{minus}{$-83$}} \\
\midrule
EM(F+P) & 194 & 35 & {\footnotesize \textcolor{minus}{$-49$}} & 57 & {\footnotesize \textcolor{minus}{$-51$}} & 65 & {\footnotesize \textcolor{minus}{$-49$}} & 82 & {\footnotesize \textcolor{minus}{$-37$}} \\
PM$^*$(F+P) & 194 & 53 & {\footnotesize \textcolor{plus}{$+4$}} & 76 & {\footnotesize \textcolor{plus}{$+2$}} & 90 & {\footnotesize \textcolor{plus}{$+4$}} & 107 & {\footnotesize \textcolor{minus}{$-3$}} \\
PM$^+$(F+P) & 194 & 66 & {\footnotesize \textcolor{plus}{$+33$}} & 96 & {\footnotesize \textcolor{plus}{$+39$}} & 119 & {\footnotesize \textcolor{plus}{$+51$}} & 138 & {\footnotesize \textcolor{plus}{$+54$}} \\
LR(F+P) & 194 & 96 & {\footnotesize \textcolor{plus}{$+8$}} & 122 & {\footnotesize \textcolor{plus}{$+2$}} & 133 & {\footnotesize \textcolor{plus}{$+4$}} & 145 & {\footnotesize \textcolor{plus}{$+6$}} \\
MLP(F+P) & 194 & 95 & {\footnotesize \textcolor{minus}{$-18$}} & 125 & {\footnotesize \textcolor{minus}{$-20$}} & 132 & {\footnotesize \textcolor{minus}{$-27$}} & 150 & {\footnotesize \textcolor{minus}{$-31$}} \\
PC(F+P) & 194 & 93 & {\footnotesize \textcolor{minus}{$-3$}} & 132 & {\footnotesize \textcolor{minus}{$-11$}} & 137 & {\footnotesize \textcolor{minus}{$-20$}} & 148 & {\footnotesize \textcolor{minus}{$-30$}} \\
\bottomrule
      \end{tabular}
    }
\end{table*}

\begin{table*}[!ht]
  \centering  
  \caption{Effectiveness of \name models on KMP scenario, using relevant tests.
  \label{tab:seshat_relevant_minimized}}  
  \scalebox{1.0}{
    \begin{tabular}{l|l|rr|rr|rr|rr}
      \toprule
      Model & Total   & \multicolumn{2}{c|}{$acc@1$} & \multicolumn{2}{c|}{$acc@3$} & \multicolumn{2}{c|}{$acc@5$} & \multicolumn{2}{c}{$acc@10$} \\
      \midrule
      EM(F+P) & 194 & 39 & {\footnotesize \textcolor{minus}{$-45$}} & 64 & {\footnotesize \textcolor{minus}{$-44$}} & 77 & {\footnotesize \textcolor{minus}{$-37$}} & 94 & {\footnotesize \textcolor{minus}{$-25$}} \\
PM$^*$(F+P) & 194 & 61 & {\footnotesize \textcolor{plus}{$+12$}} & 91 & {\footnotesize \textcolor{plus}{$+17$}} & 108 & {\footnotesize \textcolor{plus}{$+22$}} & 131 & {\footnotesize \textcolor{plus}{$+21$}} \\
PM$^+$(F+P) & 194 & 66 & {\footnotesize \textcolor{plus}{$+33$}} & 99 & {\footnotesize \textcolor{plus}{$+42$}} & 114 & {\footnotesize \textcolor{plus}{$+46$}} & 139 & {\footnotesize \textcolor{plus}{$+55$}} \\
LR(F+P) & 194 & 63 & {\footnotesize \textcolor{minus}{$-25$}} & 94 & {\footnotesize \textcolor{minus}{$-26$}} & 112 & {\footnotesize \textcolor{minus}{$-17$}} & 128 & {\footnotesize \textcolor{minus}{$-11$}} \\
MLP(F+P) & 194 & 81 & {\footnotesize \textcolor{minus}{$-32$}} & 109 & {\footnotesize \textcolor{minus}{$-36$}} & 121 & {\footnotesize \textcolor{minus}{$-38$}} & 143 & {\footnotesize \textcolor{minus}{$-38$}} \\
PC(F+P) & 194 & 71 & {\footnotesize \textcolor{minus}{$-25$}} & 109 & {\footnotesize \textcolor{minus}{$-34$}} & 121 & {\footnotesize \textcolor{minus}{$-36$}} & 137 & {\footnotesize \textcolor{minus}{$-41$}} \\
\bottomrule
      \end{tabular}
    }
\end{table*}

\subsection{Localisation Effectiveness on KMP (RQ3)}
\label{sec:result_seshat}

 We compute the scores of \name models by following the KMP scenario that
  uses a predictive kill matrix by \se. Table~\ref{tab:seshat_minimized} shows the aggregated numbers of each $acc@n$
  metric as well as the difference with the results of the FCE scenario shown in
  Table~\ref{tab:RQ1_total_ranks}. If the KMP scenario was better than the FCE scenario,
  we mark the differences with a plus sign, and if not, we mark them with a minus
  sign. 

 The most noticeable results in Table~\ref{tab:seshat_minimized} is that, in
  case of some models, \name with predicted kill matrices outperforms \name with
  actual kill matrices. This seemingly counter-intuitive result is due to the
  way inaccurate predictions of \se affects \name. Suppose \se predicts the kill
  matrix with the bias of more kills: this may result in an increased number of
  killed mutants in the faulty method, increasing its ranking score in turn. On
  the other hand, if \se predicts the kill matrix with the bias of fewer kills,
  it will result in the decreased number of killed mutants in the non-faulty
  methods, having a similar effect as that of passing test reducing the
  suspiciousness of the parts of the program they cover. Notably, this effect
  occurs strongly in partial models. In contrast, the performance of EM model
  degrades significantly. This is because, if \name is to perform well, \se has
  to make a perfectly accurate prediction for the mutants in both the faulty
  method (to match F) and the non-faulty methods (to match P). Interestingly,
  the performance improvement from \se only applies to F+P models, suggesting
  that the support from P matches has a significant effect on the performance of
  \name, as confirmed in RQ1. It is worth noting that MLP(F+P) model could not
  exhibit the best performance in terms of $acc@1$ metric, localising 18 fewer faults
  compared to the FCE scenario. This may be due to the superior capability of MLP model
  that can successfully learn the wrong predictions in the predicted kill matrix,
  which can result in a decrease in $acc@1$.

 Note that, in RQ1, \name uses only relevant tests as defined by \dfj. However, the
  results in Table~\ref{tab:seshat_minimized} is produced by predicting the
  entire kill matrices including all tests, meaning that \name with \se has more information to localise
  faults. To investigate the impact of this difference, we also conducted the
  experiment for RQ3 by only predicting the kill matrix for relevant test cases
  using \se, the results of which are listed in
  Table~\ref{tab:seshat_relevant_minimized}. We can only use F+P
  models for this experiment, as the set of relevant tests includes both failing
  and passing tests. The $acc@n$ metrics produced with relevant tests are slightly
  degraded, but the overall trend remains the same: the performance of partial
  matching models either remains the same or slightly improved, but the
  performance of others degraded. Interestingly, the performance of EM(F+P)
  model is better when using relevant tests: since there are fewer tests to
  achieve exact match, it becomes easier to score high with EM(F+P) model. 

  \begin{tcolorbox}[boxrule=0pt,frame hidden,sharp corners,enhanced,borderline north={1pt}{0pt}{black},borderline south={1pt}{0pt}{black},boxsep=2pt,left=2pt,right=2pt,top=2.5pt,bottom=2pt]
\textbf{Answer to RQ3}: Kill matrix prediction can be successfully used with
  \name to handle test cases that are introduced after the reference version. 
  \end{tcolorbox}

  \begin{table*}[!ht]
    \centering
    \caption{Uniform and stratified random sampling\label{tab:RQ3}}
    \scalebox{0.9}{
    \begin{tabular}{l|lr|rrrr||l|rrrr}
    \toprule
    Ratio & Model & Total & \multicolumn{4}{c||}{$acc$} & N & \multicolumn{4}{c}{$acc$} \\
    &   & Studied & @1 & @3 & @5 & @10 & (Ratio) & @1 & @3 & @5 & @10 \\
    \midrule
    \multirow{11}{*}{\shortstack[l]{0.1}} & EM(F) & 357 & 59.80 & 95.35 & 111.05 & 136.70 & \multirow{11}{*}{\shortstack[l]{5 \\ (0.27)}} & 36.50 & 65.05 & 83.70 & 116.30 \\
    & PM$^*$(F) & 357 & 66.55 & 107.85 & 126.55 & 159.35 & & 40.95 & 74.75 & 94.95 & 133.20 \\
    & PM$^+$(F) & 357 & 68.40 & 108.50 & 128.50 & 162.25 &  & 40.05 & 75.00 & 97.25 & 134.20 \\
    & LR(F) & 357 & 71.75 & 121.50 & 146.85 & 180.20 &  & 49.15 & 91.10 & 116.80 & 156.55 \\
    & MLP(F) & 357 & 76.70 & 120.05 & 142.40 & 177.75 &  & 49.90 & 88.85 & 115.40 & 151.05 \\
    & EM(F+P) & 224 & 46.35 & 56.55 & 62.10 & 70.75 &  & 39.50 & 57.05 & 64.20 & 69.10 \\
    & PM$^*$(F+P) & 224 & 45.45 & 66.65 & 78.90 & 95.30 &  & 65.70 & 100.25 & 111.70 & 127.95 \\
    & PM$^+$(F+P) & 224 & 29.35 & 52.45 & 64.10 & 78.80 &  & 39.55 & 52.45 & 62.70 & 81.95 \\
    & LR(F+P) & 224 & 70.70 & 93.30 & 104.30 & 117.10 &  & 75.95 & 114.40 & 128.30 & 146.05 \\
    & MLP(F+P) & 224 & 83.60 & 111.30 & 122.10 & 139.60 &  & 78.15 & 118.20 & 134.50 & 152.00 \\
    & PC(F+P) & 224 & 82.40 & 112.30 & 123.50 & 139.60 & & 70.20 & 113.25 & 129.65 & 147.10 \\
    \midrule
    \multirow{11}{*}{\shortstack[l]{0.3}} & EM(F) & 357 & 72.25 & 118.55 & 137.40 & 172.75 &\multirow{11}{*}{\shortstack[l]{10 \\ (0.41)}} & 45.60 & 80.95 & 106.15 & 145.50 \\
    & PM$^*$(F) & 357 & 78.90 & 132.15 & 156.25 & 196.20 &  & 49.65 & 93.50 & 122.15 & 161.40 \\
    & PM$^+$(F) & 357 & 83.45 & 133.70 & 159.40 & 200.40 &  & 53.65 & 93.00 & 120.20 & 162.60 \\
    & LR(F) & 357 & 84.85 & 142.15 & 166.65 & 204.90 &  & 59.95 & 106.60 & 136.50 & 179.05 \\
    & MLP(F) & 357 & 82.45 & 139.40 & 164.10 & 202.50 &  & 56.90 & 102.60 & 132.55 & 174.10 \\
    & EM(F+P) & 224 & 66.70 & 82.35 & 89.25 & 95.40 &  & 50.60 & 74.00 & 80.85 & 85.50 \\
    & PM$^*$(F+P) & 224 & 47.80 & 71.40 & 83.35 & 104.75 &  & 74.45 & 106.35 & 120.15 & 140.65 \\
    & PM$^+$(F+P) & 224 & 31.75 & 55.45 & 67.70 & 82.75 & & 39.20 & 55.05 & 67.20 & 81.65 \\
    & LR(F+P) & 224 & 81.95 & 107.60 & 116.40 & 131.50 & & 82.15 & 115.90 & 128.80 & 146.70 \\
    & MLP(F+P) & 224 & 101.55 & 132.20 & 143.40 & 159.85 & & 89.70 & 126.35 & 141.60 & 161.45 \\
    & PC(F+P) & 224 & 95.20 & 133.30 & 144.95 & 162.90 & & 84.15 & 127.90 & 144.40 & 163.25 \\
    \midrule
    \multirow{11}{*}{\shortstack[l]{0.5}} & EM(F) & 357 & 75.90 & 128.30 & 149.35 & 188.55 &\multirow{11}{*}{\shortstack[l]{15 \\ (0.50)}} & 53.60 & 96.70 & 122.20 & 160.65 \\
    & PM$^*$(F) & 357 & 82.90 & 141.80 & 168.85 & 211.15 && 58.70 & 110.10 & 139.65 & 177.95 \\
    & PM$^+$(F) & 357 & 86.30 & 143.80 & 173.60 & 215.20 && 62.20 & 109.15 & 142.75 & 182.70 \\
    & LR(F) & 357 & 88.75 & 145.90 & 175.75 & 216.65 && 66.70 & 118.30 & 152.35 & 190.90 \\
    & MLP(F) & 357 & 84.70 & 146.00 & 170.95 & 212.80 && 63.05 & 114.25 & 146.00 & 187.10 \\
    & EM(F+P) & 224 & 73.90 & 93.10 & 100.05 & 106.50 && 57.35 & 82.65 & 89.45 & 94.75 \\
    & PM$^*$(F+P) & 224 & 47.55 & 72.70 & 84.20 & 108.00 && 78.75 & 109.90 & 123.30 & 141.50 \\
    & PM$^+$(F+P) & 224 & 32.70 & 57.00 & \textbf{68.65} & 83.30 && \textbf{42.15} & 61.45 & \textbf{71.60} & 82.15 \\
    & LR(F+P) & 224 & 86.35 & 112.15 & 121.75 & 135.20 && 86.95 & 116.95 & 129.70 & 148.35 \\
    & MLP(F+P) & 224 & 104.90 & 138.65 & 151.35 & 166.70 && 94.80 & 134.60 & 147.95 & 163.60 \\
    & PC(F+P) & 224 & 98.10 & 138.85 & 151.80 & 170.05 & & 91.85 & 133.85 & 151.30 & 169.95 \\
    \midrule
    \multirow{11}{*}{\shortstack[l]{0.7}} & EM(F) & 357 & \textbf{78.05} & 133.55 & 156.30 & 194.60 & \multirow{11}{*}{\shortstack[l]{20 \\ (0.56)}} & 55.80 & 105.80 & 133.25 & 172.90 \\
    & PM$^*$(F) & 357 & \textbf{84.80} & 147.75 & 175.60 & 216.90 && 64.95 & 121.45 & 153.75 & 193.75 \\
    & PM$^+$(F) & 357 & \textbf{88.15} & 150.05 & 177.85 & 219.30 && 70.05 & 124.10 & 156.30 & 195.10 \\
    & LR(F) & 357 & 89.65 & 148.50 & 179.85 & 221.25 && 74.35 & 127.05 & 162.95 & 200.15 \\
    & MLP(F) & 357 & 84.30 & 144.60 & 173.40 & 214.85 && 69.70 & 124.70 & 158.15 & 196.60 \\
    & EM(F+P) & 224 & 78.05 & 98.85 & 106.05 & 112.15 && 62.50 & 88.05 & 94.05 & 100.55 \\
    & PM$^*$(F+P) & 224 & 48.30 & 73.60 & 83.95 & 108.35 && \textbf{81.55} & \textbf{110.15} & \textbf{125.50} & \textbf{142.60} \\
    & PM$^+$(F+P) & 224 & 32.90 & \textbf{57.20} & 68.10 & 83.40 && 40.55 & \textbf{62.40} & 69.05 & 81.65 \\
    & LR(F+P) & 224 & 87.20 & 115.20 & 125.15 & 137.85 && \textbf{90.70} & 118.60 & \textbf{131.60} & \textbf{148.90} \\
    & MLP(F+P) & 224 & 108.45 & 142.55 & 155.30 & 170.20 && 98.25 & 138.20 & 150.00 & 165.90 \\
    & PC(F+P) & 224 & \textbf{99.75} & 142.30 & 155.30 & 174.35 & & \textbf{97.25} & 137.75 & 152.75 & 172.95 \\
    \midrule
    \multirow{11}{*}{\shortstack[l]{Full}} & EM(F) & 357 & 77.00 & \textbf{139.00} & \textbf{164.00} & \textbf{204.00} & \multirow{11}{*}{\shortstack[l]{Full}} & \textbf{77.00} & \textbf{139.00} & \textbf{164.00} & \textbf{204.00} \\
    & PM$^*$(F) & 357 & 82.00 & \textbf{151.00} & \textbf{183.00} & \textbf{224.00} && \textbf{82.00} & \textbf{151.00} & \textbf{183.00} & \textbf{224.00} \\
    & PM$^+$(F) & 357 & 86.00 & \textbf{155.00} & \textbf{186.00} & \textbf{227.00} && \textbf{86.00} & \textbf{155.00} & \textbf{186.00} & \textbf{227.00} \\
    & LR(F) & 357 & \textbf{90.00} & \textbf{152.00} & \textbf{186.00} & \textbf{226.00} && \textbf{90.00} & \textbf{152.00} & \textbf{186.00} & \textbf{226.00} \\
    & MLP(F) & 357 & \textbf{85.00} & \textbf{150.00} & \textbf{180.00} & \textbf{223.00} && \textbf{85.00} & \textbf{150.00} & \textbf{180.00} & \textbf{223.00} \\
    & EM(F+P) & 224 & \textbf{84.00} & \textbf{108.00} & \textbf{114.00} & \textbf{119.00} && \textbf{84.00} & \textbf{108.00} & \textbf{114.00} & \textbf{119.00} \\
    & PM$^*$(F+P) & 224 & \textbf{49.00} & \textbf{74.00} & \textbf{86.00} & \textbf{110.00} && 49.00 & 74.00 & 86.00 & 110.00 \\
    & PM$^+$(F+P) & 224 & \textbf{33.00} & 57.00 & 68.00 & \textbf{84.00} && 33.00 & 57.00 & 68.00 & \textbf{84.00} \\
    & LR(F+P) & 224 & \textbf{88.00} & \textbf{120.00} & \textbf{129.00} & \textbf{139.00} && 88.00 & \textbf{120.00} & 129.00 & 139.00 \\
    & MLP(F+P) & 224 & \textbf{113.00} & \textbf{145.00} & \textbf{159.00} & \textbf{181.00} && \textbf{113.00} & \textbf{145.00} & \textbf{159.00} & \textbf{181.00} \\  
    & PC(F+P) & 224 & 96.00 & \textbf{143.00} & \textbf{157.00}  & \textbf{178.00} & & 96.00 & \textbf{143.00} & \textbf{157.00}  & \textbf{178.00}
     \\
    \bottomrule
    \end{tabular}
    }
  \end{table*}

\begin{figure}[ht]
  \centering      
  \includegraphics[width=.45\textwidth]{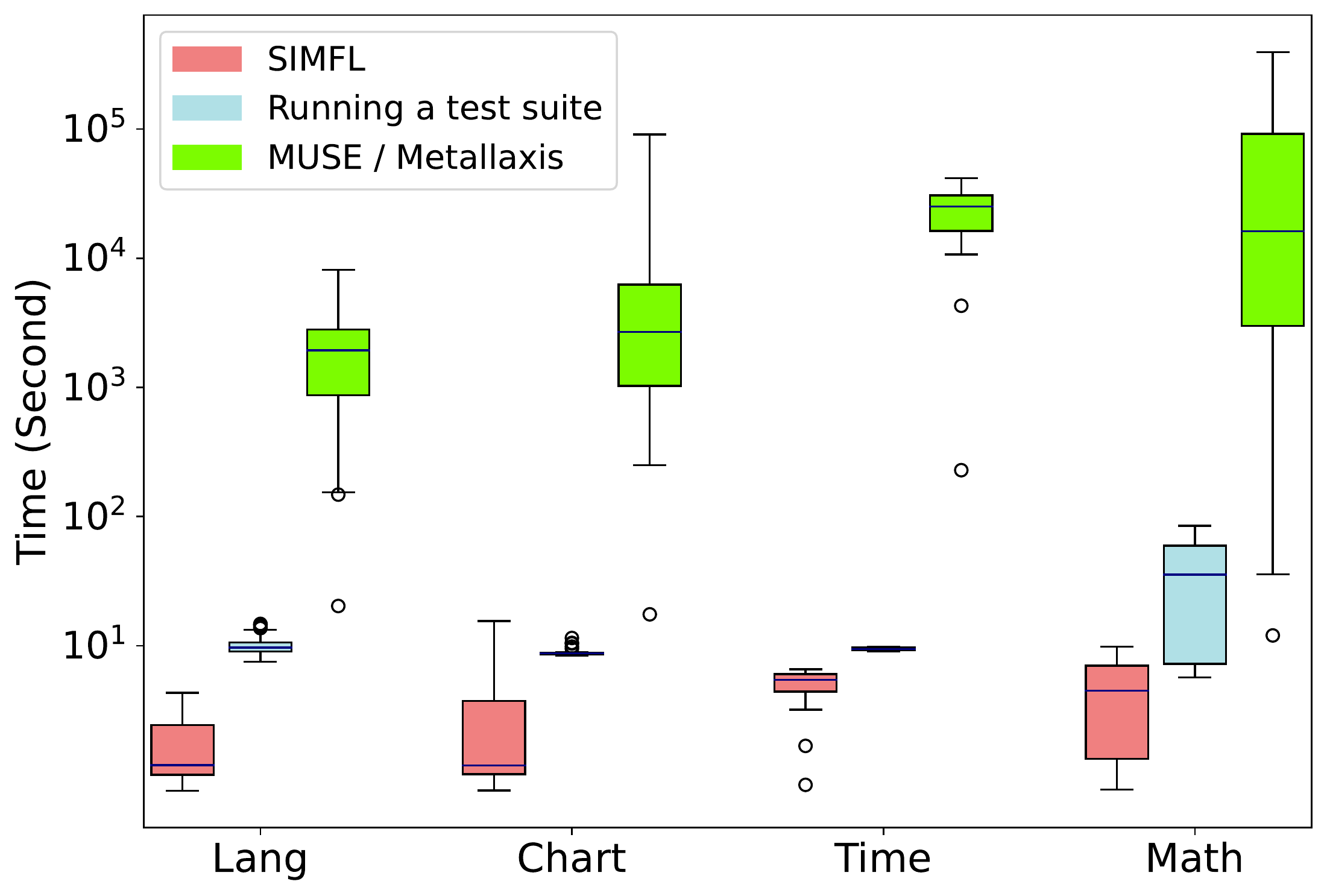}
  \caption{Execution time of \name compared to MUSE, Metallaxis, and a full
  test suite run. Y-axis is shown on a logarithmic scale. \label{fig:RQX_time_comp}}
\end{figure}

\subsection{Efficiency (RQ4)}
\label{sec:time_comparison}
\name is different from the other MBFL techniques in that it does not require
test runs against the generated mutants after the fault is observed. Therefore,
it may have an advantage over the existing MBFL techniques in terms of execution
time and a faster feedback to the developers. To quantify this advantage, we
compare the execution time of \name with that of MUSE and Metallaxis, and a full
test suite run. Note that we compare the online cost of \name with other
MBFL techniques that do not require generating a kill matrix in advance and
hence have no offline cost. See Section~\ref{sec:threats} for the discussion
on the online and offline costs of \name.


To measure the execution time of \name, we use the MLP model as it is the
slowest model due to the training and inference time of it. As MUSE
and Metallaxis lack open-source implementation, we employ a surrogate estimator
to measure their execution time. That is, for each subject program, we collect the
mutants covered by failing tests and measure the time for running the tests that
cover them. Although MUSE and Metallaxis employ different formulae for
calculating suspiciousness scores, we approximate their execution time to be
the same because both require the same test runs to compute the scores (i.e., whether
the mutants are killed or not) and the other computational costs are negligible compared
to the test runs.

Figure~\ref{fig:RQX_time_comp} illustrates the execution time of running
each technique, broken down by each project. Overall, \name is on average 4.7x
faster than running a full test suite and notably faster than MUSE and
Metallaxis, showing on average 5,687x faster execution. The time gap between
\name and MUSE and Metallaxis is even larger for larger projects: \name is on
average 1,226x faster than MUSE and Metallaxis for the faults in Lang but
13,366x faster for the faults in Math. This result highlights the positive
affect of the cost amortisation of \name.

\begin{tcolorbox}[boxrule=0pt,frame hidden,sharp corners,enhanced,borderline north={1pt}{0pt}{black},borderline south={1pt}{0pt}{black},boxsep=2pt,left=2pt,right=2pt,top=2.5pt,bottom=2pt]
\textbf{Answer to RQ4}: \name is considerably faster than both MUSE
and Metallaxis and more efficient than running a full test suite. These results
showcase the efficient fault localization capabilities of \name, despite using mutants.
\end{tcolorbox}

\subsection{Effects of Sampling (RQ5)}
\label{sec:result_RQ3}


To investigate how the mutation sampling rates affect the performance of \name,
we attempt to localise the studied faults using mutants sampled with different
rates. Table~\ref{tab:RQ3} (left side) shows the uniform sampling results with  
rates of 0.1, 0.3, 0.5, and 0.7: all metric values are 
averaged across 20 different samples.
Table~\ref{tab:RQ3} also 
includes the results obtained without sampling (Full). The best results are 
typeset in bold.

As expected, the Full configuration often shows the best performance, followed
by sampling rates of 0.7 and 0.5. Since we expect different mutants to 
contribute different amounts of information to localisation, we do not find it
surprising that sampling rates down to 0.5 show comparable results with
the Full configuration. However, the performance does not degrade at the same
rate as the sampling rate, as can be seen from the results obtained using 
the sampling rate of 0.1.

Since larger methods are likely to produce more mutants, uniform sampling will
effectively sample more mutants for larger methods. We investigate whether 
this is disadvantageous for relatively smaller methods by evaluating stratified
sampling: given the threshold parameter $N$, stratified sampling randomly 
chooses only $N$ mutants from methods with more than $N$ mutants, and chooses
all mutants if their number is below $N$. Table~\ref{tab:RQ3} (right side)
contains the results obtained using stratified mutant sampling with $N \in \{5, 10, 15, 20\}$.
The value in the parenthesis, i.e., ``Ratio'', is the average ratio of the 
number of mutants sampled by stratified sampling to the number of all mutants.

Compared to the Full configuration, the performance degradation as $N$ 
decreases is notably worse than what has been observed from the results of 
uniform random sampling. However, even with $N = 5$, the sample ratio is 
0.27 on average, higher than the smallest sampling rate for the uniform 
sampling. The comparison suggests that, contrary to our concern for a 
potential bias against smaller methods, stratified sampling is actually 
harmful to \name. One interpretation of the result is that, if we assume that
the location of a fault is a random variable, larger methods are
by definition more likely to contain it.

\begin{tcolorbox}[boxrule=0pt,frame hidden,sharp corners,enhanced,borderline north={1pt}{0pt}{black},borderline south={1pt}{0pt}{black},boxsep=2pt,left=2pt,right=2pt,top=2.5pt,bottom=2pt]
\textbf{Answer to RQ5}: The impact of mutation sampling is observable but not 
too disruptive. Using uniform sampling, on average 80\%
of the faults ranked at the first rank without sampling can still be localised
at the first rank. However, stratified sampling actually harms \name: larger 
methods need to be represented by more mutants.
\end{tcolorbox}

\begin{table*}[!ht]
  \centering  
  \caption{ The effects of removing subsumed mutants. For the results of each
  project, see Table~\ref{tab:dominance_extended}.
  \label{tab:dominance_minimized}}  
  \scalebox{1.0}{
  \begin{tabular}{l|l|rr|rr|rr|rr}
    \toprule
    Model & Total   & \multicolumn{2}{c|}{$acc@1$} & \multicolumn{2}{c|}{$acc@3$} & \multicolumn{2}{c|}{$acc@5$} & \multicolumn{2}{c}{$acc@10$} \\
    \midrule
    EM(F) & 357 & 92 & {\footnotesize \textcolor{plus}{$+15$}} & 129 & {\footnotesize \textcolor{minus}{$-10$}} & 144 & {\footnotesize \textcolor{minus}{$-19$}} & 164 & {\footnotesize \textcolor{minus}{$-40$}} \\
  PM$^*$(F) & 357 & 126 & {\footnotesize \textcolor{plus}{$+44$}} & 173 & {\footnotesize \textcolor{plus}{$+22$}} & 191 & {\footnotesize \textcolor{plus}{$+8$}} & 211 & {\footnotesize \textcolor{minus}{$-14$}} \\
  PM$^+$(F) & 357 & 123 & {\footnotesize \textcolor{plus}{$+37$}} & 170 & {\footnotesize \textcolor{plus}{$+15$}} & 189 & {\footnotesize \textcolor{plus}{$+3$}} & 208 & {\footnotesize \textcolor{minus}{$-19$}} \\
  LR(F) & 357 & 122 & {\footnotesize \textcolor{plus}{$+32$}} & 170 & {\footnotesize \textcolor{plus}{$+18$}} & 184 & {\footnotesize \textcolor{minus}{$-2$}} & 216 & {\footnotesize \textcolor{minus}{$-10$}} \\
  MLP(F) & 357 & 136 & {\footnotesize \textcolor{plus}{$+51$}} & 183 & {\footnotesize \textcolor{plus}{$+33$}} & 206 & {\footnotesize \textcolor{plus}{$+26$}} & 230 & {\footnotesize \textcolor{plus}{$+7$}} \\
  \midrule
  EM(F+P) & 224 & 75 & {\footnotesize \textcolor{minus}{$-9$}} & 96 & {\footnotesize \textcolor{minus}{$-12$}} & 106 & {\footnotesize \textcolor{minus}{$-8$}} & 115 & {\footnotesize \textcolor{minus}{$-4$}} \\
  PM$^*$(F+P) & 224 & 71 & {\footnotesize \textcolor{plus}{$+22$}} & 110 & {\footnotesize \textcolor{plus}{$+36$}} & 123 & {\footnotesize \textcolor{plus}{$+37$}} & 143 & {\footnotesize \textcolor{plus}{$+33$}} \\
  PM$^+$(F+P) & 224 & 44 & {\footnotesize \textcolor{plus}{$+11$}} & 91 & {\footnotesize \textcolor{plus}{$+34$}} & 107 & {\footnotesize \textcolor{plus}{$+39$}} & 139 & {\footnotesize \textcolor{plus}{$+55$}} \\
  LR(F+P) & 224 & 103 & {\footnotesize \textcolor{plus}{$+15$}} & 134 & {\footnotesize \textcolor{plus}{$+14$}} & 145 & {\footnotesize \textcolor{plus}{$+16$}} & 156 & {\footnotesize \textcolor{plus}{$+17$}} \\
  MLP(F+P) & 224 & 119 & {\footnotesize \textcolor{plus}{$+6$}} & 150 & {\footnotesize \textcolor{plus}{$+5$}} & 157 & {\footnotesize \textcolor{minus}{$-2$}} & 168 & {\footnotesize \textcolor{minus}{$-13$}} \\
  PC(F+P) & 224 & 102 & {\footnotesize \textcolor{plus}{$+6$}} & 134 & {\footnotesize \textcolor{minus}{$-9$}} & 145 & {\footnotesize \textcolor{minus}{$-12$}} & 155 & {\footnotesize \textcolor{minus}{$-23$}} \\
    \bottomrule
    \end{tabular}
  }
\end{table*} 

\subsection{Effects of Subsumed Mutants (RQ6)}
\label{sec:results_RQ6}

It has been pointed out that mutation tools generate numerous subsumed and
redundant mutants, resulting in a mutation score to be easily
misunderstood~\cite{just2012redundant, ammann2014establishing}. As the killing
of the subsuming mutants always accompany the killing of their subsumed mutants,
this problem not only inflates the mutation score but also inflate our
method-level FL scores by \name models. We investigate the effects of
removing subsumed mutants. First, we identify the most subsuming
mutants~\cite{papadakis2016threats}, which are located at the first rank of DMSG
(Dynamic Mutant Subsumption Graph)~\cite{ammann2014establishing} and are not
subsumed by other mutants except their indistinguishable mutants.\footnote{The
two mutants are said to be indistinguishable if they have the same test results for
all tests.} Next, we conduct RQ1 study using only subsuming mutants and compare
$acc@n$ with the results of the original RQ1 study that uses all mutants.

As shown in Table~\ref{tab:dominance_minimized}, $acc@1$ of all models
  except EM(F+P) largely improved: MLP(F) localised 51 more faults at the first rank
  compared to the model using all mutants. The reason EM(F+P) worsened for all
  $acc@n$ is that the subsuming mutants are likely to be killed by a few tests
  and they eliminate subsumed mutants needed for exact matching. This reason
  also explains why the overall $acc@n$ of F models improved more than F+P
  models.

\begin{tcolorbox}[boxrule=0pt,frame hidden,sharp corners,enhanced,borderline north={1pt}{0pt}{black},borderline south={1pt}{0pt}{black},boxsep=2pt,left=2pt,right=2pt,top=2.5pt,bottom=2pt]
\textbf{Answer to RQ6}: Although we used fewer mutants,
removing subsumed mutants brings positive effects to \name models by increasing
the number of faults located at the first rank up to 51, compared to using all mutants.
\end{tcolorbox}

\section{Discussion}
\label{sec:discussion}

\subsection{Relation with Other FL Techniques}
\label{sec:relation}

\begin{figure}[ht]
  \centering
  \begin{minipage}[t]{.22\textwidth}
    \centering
        \includegraphics[width=\textwidth]{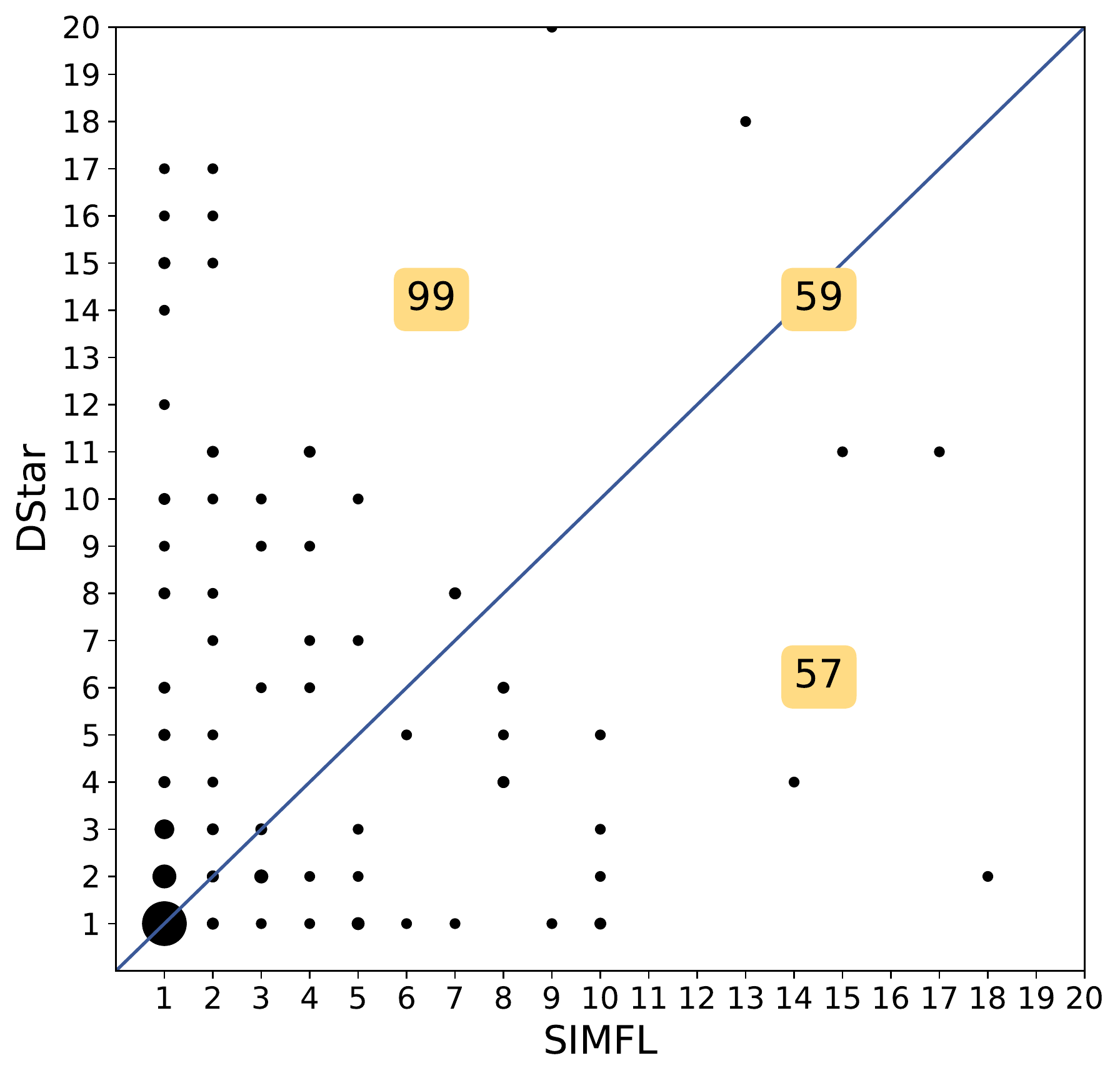}
        \vspace{-1.5em}
        \subcaption{DStar ($r = 0.218$)\label{fig:vs_dstar}}
  \end{minipage}%
  \begin{minipage}[t]{.22\textwidth}
    \centering
        \includegraphics[width=\textwidth]{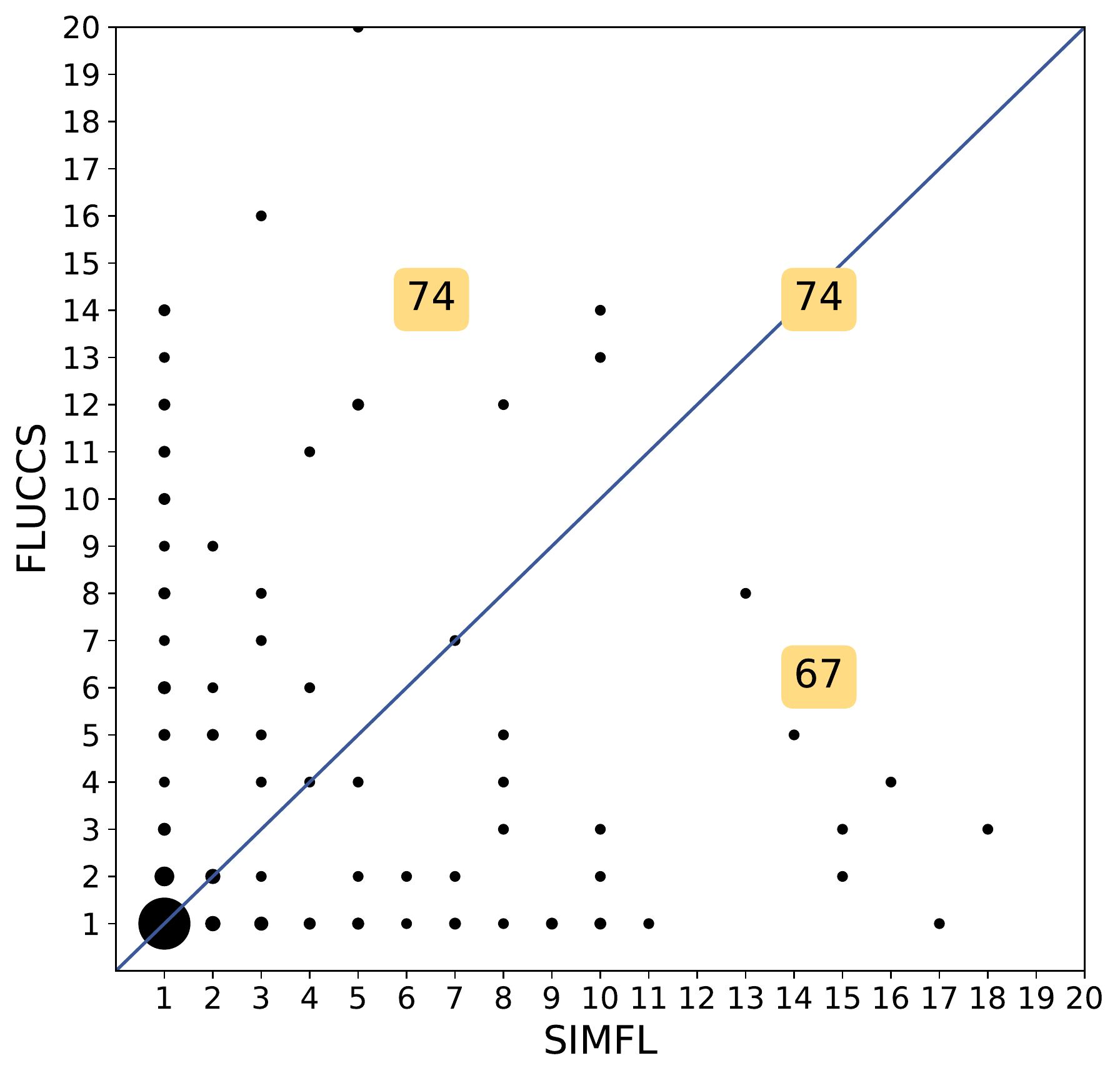}
        \vspace{-1.5em}
        \subcaption{FLUCCS ($r = 0.105$)\label{fig:vs_fluccs}}
  \end{minipage}
  \begin{minipage}[t]{.22\textwidth}
      \centering
      \includegraphics[width=\textwidth]{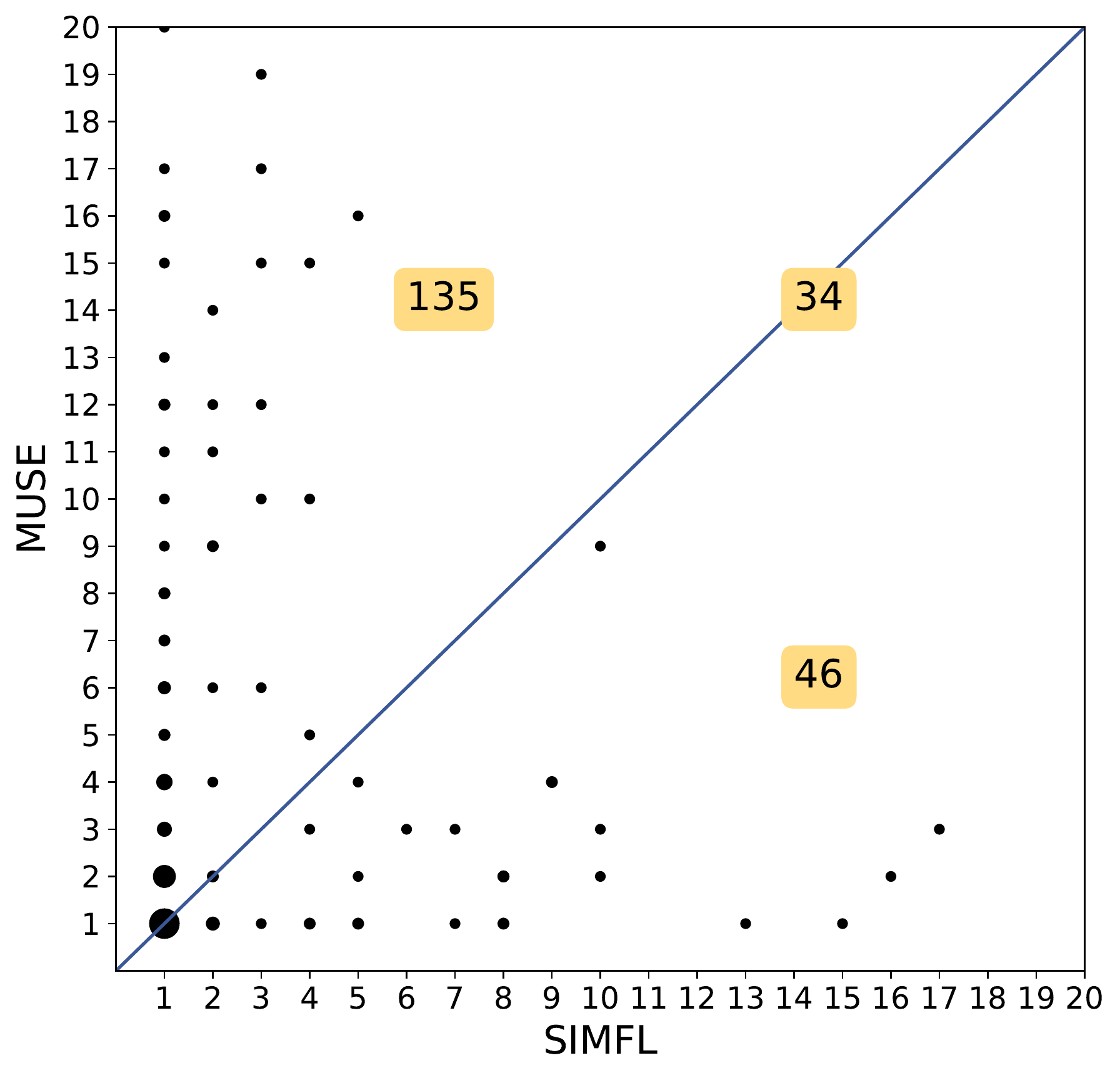}
      \vspace{-1.5em}
      \subcaption{MUSE ($r = 0.072$)\label{fig:vs_muse}}
  \end{minipage}%
  \begin{minipage}[t]{.22\textwidth}
    \centering
      \includegraphics[width=\textwidth]{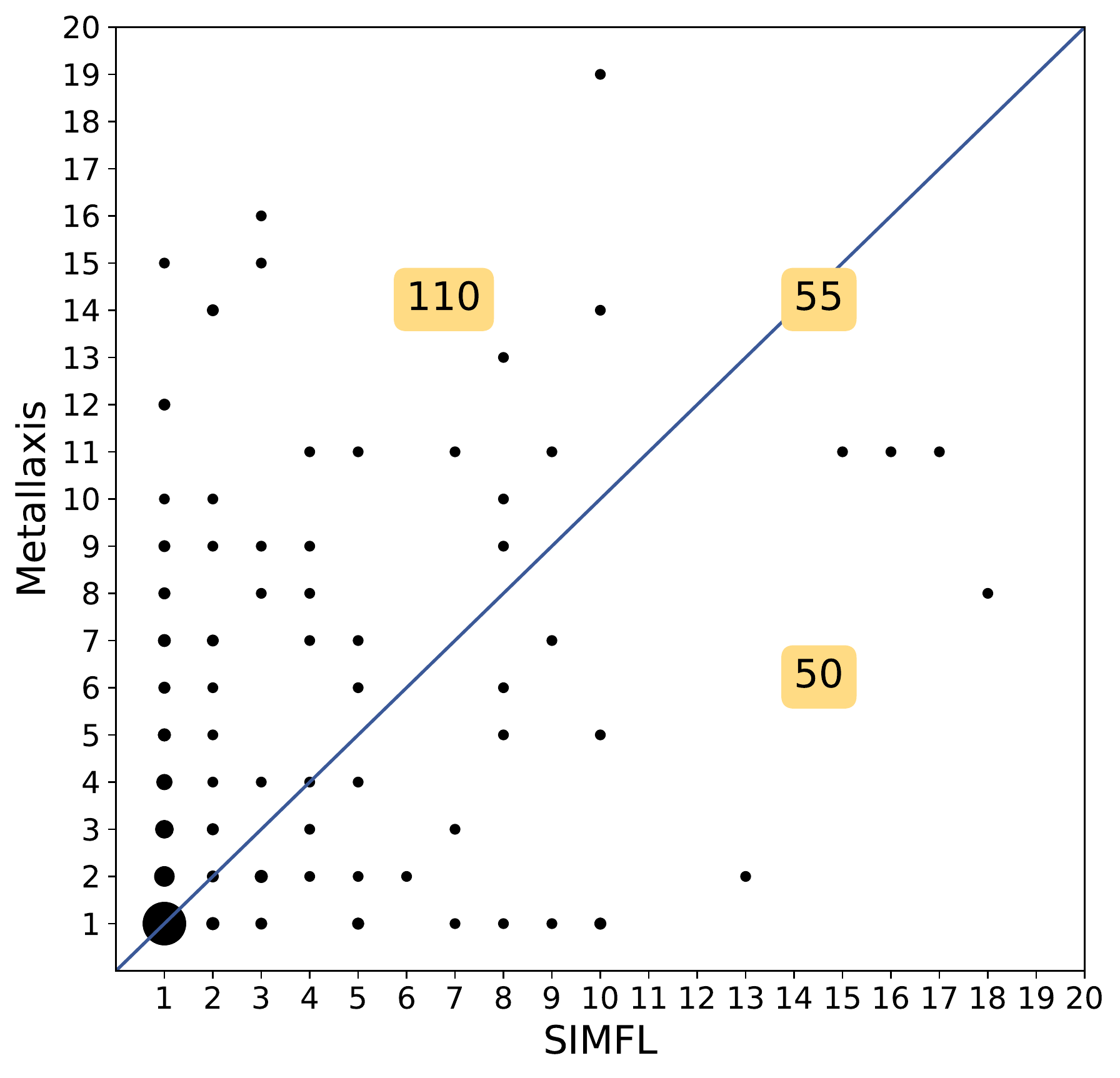}
      \vspace{-1.5em}
      \subcaption{Metallaxis ($r = 0.130$)\label{fig:vs_metallaxis}}
  \end{minipage}
  \caption{Comparison of MLP(F+P) and other FL techniques.
  \label{fig:one_to_one_comparison}}
\end{figure}

Two FL techniques can be complementary to each other if there is little overlap
between faults ranked highly by each technique. 
To investigate whether the contribution of \name is uniquely different from 
others, we investigate how individual faults are ranked differently by \name 
and other FL techniques. \name is represented by the MLP(F+P) model. 
We omit TraPT from this comparison as the individual 
rank information was not available from the paper.

Figure~\ref{fig:one_to_one_comparison} plots each individual fault according 
to its rank by MLP(F+P) of \name ($x$-axis) and its rank by the other FL 
technique ($y$-axis). Data points on the line $y = x$ represent faults that 
are ranked at the same place by both techniques, whereas the farther from the 
line a point is, the more differently it is ranked by two techniques. Plots
only contain faults that are ranked within the top 20 places by at least one
technique: the size of the dots corresponds to the number of faults plotted
at the location of the dot. The numbers on the $y=x$ line as well as above and 
below the line show the total number of faults that belong to the corresponding 
parts, regardless of being ranked within the top 20 or not. For example, \name 
ranks 135 faults higher than MUSE.
The agreement between two techniques are measured using Pearson correlation 
coefficient ($r$): value 0 implies no correlation and, therefore, no 
agreement, whereas value 1 implies perfect correlation and, therefore, two 
identical rankings.

While there exist dense clusters of points near the top ranks around the 
$y=x$ line, there is no clear relationship between FL techniques. \name
shows low Pearson correlation coefficients against all compared techniques.
Notably, \name is significantly different from two existing MBFL techniques,
MUSE and Metallaxis, suggesting that the way \name captures the relationship
between faults and tests differs significantly from existing MBFL techniques.
\name also ranks the most faults identically to FLUCCS, a technique that uses
multiple SBFL scores as well as code and change metric, suggesting that 
mutation analysis can be a rich source of information for fault localisation.
Overall, the results provide evidence that there exist 
faults that \name can localise much more effectively than the other, and vice 
versa. The complementary nature also suggests the possibility of an effective 
hybridisation of \name and other techniques, as recent work that combine
multiple FL techniques suggest~\cite{zou2019empirical,Xuan:2014kq,Sohn:2017xq}. 
We leave the hybridisation as future work.


\subsection{Test Case Granularity}
\label{sec:granularity}

 
\begin{table*}[ht]
  \centering  
  \caption{The result of Mann–Whitney U test on the number of methods whose
  mutants are killed by the failing test cases, between the faults localised within ($W3$)
  and out of top three places ($O3$). The significant level is 0.05: if $p$-value is less then 0.05 (typeset in bold),
  then we can \emph{reject} null hypothesis that there is no significant difference between the two groups from which the samples were drawn.
  \label{tab:test_gran}}
  \scalebox{0.8}{
    \begin{tabular}{l|l|rrr|rrr|r||l|l|rrr|rrr|r}
      \toprule
      Model & Project & \multicolumn{3}{c|}{$W3$} & \multicolumn{3}{c|}{$O3$} & $p$ & Model & Project & \multicolumn{3}{c|}{$W3$} & \multicolumn{3}{c|}{$O3$} & $p$ \\
            &         & mean & std & sp                       & mean & std   & sp                         &     &       &         & mean & std & sp & mean & std      & sp                                             & \\
      \midrule
      \multirow{5}{*}{\shortstack[l]{EM\\(F)}} & Lang & 5.5 & 5.2 & 45 & 10.0 & 12.9 & 17 & 0.171 & \multirow{5}{*}{\shortstack[l]{EM\\(F+P)}} & Lang & 6.2 & 5.7 & 41 & 7.9 & 12.0 & 20 & 0.945 \\
      & Chart & 10.0 & 6.8 & 11 & 144.9 & 260.5 & 15 & \textbf{0.002} &  & Chart & 11.4 & 7.3 & 9 & 136.1 & 254.5 & 16 & \textbf{0.017} \\
      & Time & 69.2 & 36.1 & 9 & 129.6 & 46.7 & 17 & \textbf{0.003} &  & Time & 103.7 & 54.2 & 13 & 113.7 & 49.2 & 13 & 0.739 \\
      & Closure & 167.2 & 117.3 & 31 & 322.7 & 182.2 & 101 & \textbf{0.000} &  & Closure & - & -  & - & -  & -     & -    & - \\
      & Math & 15.5 & 19.1 & 43 & 34.8 & 26.6 & 59 & \textbf{0.000} &  & Math & 20.5 & 25.1 & 45 & 28.5 & 21.7 & 46 & \textbf{0.022} \\
     \midrule
     \multirow{5}{*}{\shortstack[l]{PM$^*$\\(F)}} & Lang & 5.1 & 3.8 & 47 & 11.9 & 14.3 & 15 & 0.125 & \multirow{5}{*}{\shortstack[l]{PM$^*$\\(F+P)}} & Lang & 4.7 & 3.7 & 36 & 9.8 & 11.6 & 25 & \textbf{0.032} \\
     & Chart & 10.0 & 6.8 & 11 & 144.9 & 260.5 & 15 & \textbf{0.002} &  & Chart & 96.3 & 242.1 & 9 & 88.4 & 193.4 & 16 & 0.074 \\
     & Time & 78.4 & 43.9 & 10 & 127.6 & 47.4 & 16 & \textbf{0.013} &  & Time & 20.0 & 27.6 & 3 & 120.3 & 42.4 & 23 & \textbf{0.010} \\
     & Closure & 175.7 & 113.8 & 36 & 327.6 & 184.9 & 96 & \textbf{0.000} &  & Closure & - & -  & - & -  & -     & -    & - \\
     & Math & 15.4 & 18.8 & 47 & 36.3 & 26.6 & 55 & \textbf{0.000} &  & Math & 15.2 & 19.5 & 26 & 28.2 & 24.3 & 65 & \textbf{0.002} \\
    \midrule
    \multirow{5}{*}{\shortstack[l]{PM$^+$\\(F)}} & Lang & 5.1 & 3.8 & 48 & 12.4 & 14.7 & 14 & 0.126 & \multirow{5}{*}{\shortstack[l]{PM$^+$\\(F+P)}} & Lang & 4.4 & 3.5 & 31 & 9.2 & 10.9 & 30 & \textbf{0.026} \\
     & Chart & 9.8 & 7.1 & 10 & 136.6 & 254.3 & 16 & \textbf{0.004} &  & Chart & 96.9 & 242.0 & 9 & 88.1 & 193.5 & 16 & 0.106 \\
     & Time & 78.4 & 43.9 & 10 & 127.6 & 47.4 & 16 & \textbf{0.013} &  & Time & 0.5 & 0.5 & 2 & 117.7 & 43.2 & 24 & \textbf{0.024} \\
     & Closure & 203.2 & 160.8 & 41 & 323.5 & 178.0 & 91 & \textbf{0.000} &  & Closure & - & -  & - & -  & -     & -    & - \\
     & Math & 15.5 & 19.1 & 46 & 35.8 & 26.5 & 56 & \textbf{0.000} &  & Math & 11.5 & 9.9 & 15 & 27.1 & 24.9 & 76 & \textbf{0.006} \\
    \midrule
    \multirow{5}{*}{\shortstack[l]{LR\\(F)}} & Lang & 4.9 & 3.9 & 49 & 13.5 & 14.7 & 13 & \textbf{0.019} & \multirow{5}{*}{\shortstack[l]{LR\\(F+P)}} & Lang & 5.5 & 5.0 & 49 & 12.2 & 14.8 & 12 & 0.122 \\
     & Chart & 11.8 & 6.2 & 9 & 135.9 & 254.6 & 16 & \textbf{0.019} &  & Chart & 70.3 & 200.9 & 14 & 117.9 & 223.1 & 11 & 0.118 \\
     & Time & 78.4 & 43.9 & 10 & 127.6 & 47.4 & 16 & \textbf{0.013} &  & Time & 90.8 & 56.3 & 14 & 129.6 & 36.6 & 12 & 0.105 \\
     & Closure & 193.3 & 136.7 & 37 & 322.3 & 184.0 & 95 & \textbf{0.000} &  & Closure & - & -  & - & -  & -     & -    & - \\
     & Math & 16.5 & 19.1 & 47 & 35.3 & 27.1 & 55 & \textbf{0.000} &  & Math & 17.3 & 18.1 & 43 & 31.0 & 26.3 & 48 & \textbf{0.002} \\
    \midrule
    \multirow{5}{*}{\shortstack[l]{ML\\P(F)}} & Lang & 5.4 & 5.1 & 51 & 12.9 & 15.0 & 11 & \textbf{0.042} & \multirow{5}{*}{\shortstack[l]{MLP\\(F+P)}} & Lang & 5.9 & 5.3 & 55 & 13.3 & 18.4 & 7 & 0.496 \\
     & Chart & 10.8 & 6.6 & 10 & 144.9 & 260.5 & 15 & \textbf{0.004} &  & Chart & 14.6 & 10.1 & 13 & 174.2 & 283.7 & 12 & \textbf{0.019} \\
     & Time & 62.3 & 40.0 & 10 & 137.7 & 34.7 & 16 & \textbf{0.000} &  & Time & 95.1 & 54.9 & 16 & 130.5 & 37.7 & 10 & 0.133 \\
     & Closure & 168.6 & 114.0 & 33 & 325.3 & 183.0 & 99 & \textbf{0.000} &  & Closure & - & -  & - & -  & -     & -    & - \\
     & Math & 15.6 & 18.9 & 46 & 35.8 & 26.7 & 56 & \textbf{0.000} &  & Math & 17.0 & 16.6 & 61 & 40.9 & 29.6 & 41 & \textbf{0.000} \\
    \midrule
      \multicolumn{9}{c||}{}  & \multirow{5}{*}{\shortstack[l]{PC\\(F+P)}} & Lang & 5.4 & 4.0 & 51 & 12.7 & 16.5 & 11 & 0.592 \\
      \multicolumn{9}{c||}{} &  & Chart & 76.6 & 193.5 & 16 & 105.7 & 230.1 & 10 & \textbf{0.033} \\
      \multicolumn{9}{c||}{} &  & Time & 91.4 & 43.7 & 15 & 132.3 & 53.0 & 11 & \textbf{0.029} \\
      \multicolumn{9}{c||}{} &  & Closure & - & -  & - & -  & -     & -    & - \\
        \multicolumn{9}{c||}{} &  & Math & 16.5 & 17.4 & 61 & 41.8 & 28.1 & 41 & \textbf{0.000} \\
      \bottomrule
    \end{tabular}    
  }
\end{table*}

A common pattern observed in all configurations of \name is that it performs 
the best for Commons Lang. Following Laghari and Demeyer~\cite{Laghari2018aa}, 
we hypothesise that this may be related to the test case granularity: if each 
test case kills mutants that exist in only a few methods, \name can benefit 
from this because failures of each test case will be tightly coupled with a 
few candidate locations.

To investigate the impact of test case granularity, we check whether the number
of the methods that are relevant to failures caused by highly ranked faults is
lower than the number of methods relevant to faults that are not ranked near 
the top. We define a method $m$ to be relevant to the failure of a test case $t$
if $t$ kills a mutant in $m$. A finer granularity test case $t$ is expected to
be relevant to fewer methods. We categorise faults into those ranked in the top
three places (set $W3$), and those that are not (set $O3$), and compare the number
of relevant methods between $W3$ and $O3$.

Table~\ref{tab:test_gran} reports the results of Mann–Whitney U test on
the number of relevant methods between $W3$ and $O3$. The column $sp$ shows the
number of samples for each group and column $p$ shows the $p$-value. For 36 out
of 49 cases, we accept the alternative hypothesis that there is a statistically
significant difference between the number of relevant methods between $W3$ and
$O3$. This supports our assumption that the faults in $W3$ are likely to be
revealed by test cases with finer-granularity than the faults in $O3$. The test
cases of Commons Lang have finer-granularity when compared to other subjects,
leading us to conjecture that test case granularity is why \name performs more
effectively against Lang than others. However, the results also show that \name
is not simply reflecting a one-to-one mapping between methods (mutants) and
their unit tests: failing test cases of Closure kill mutants in 203 methods on
average, but PM$^+$(F) can still localise 41 out of 132 faults within the top
three places (see Table~\ref{tab:RQ1_total_ranks}).

\subsection{Kill Reason Filtering}
\label{sec:kill_reason_filtering}
\begin{table*}[ht]
  \centering  
  \caption{
  The $acc@n$ metric values after filtering mutants based on their kill reasons.\label{tab:kill_reason_filtering}}  
  \scalebox{0.9}{
    \begin{tabular}{l|l|rrr||rrr|rrr|rrr}
      \toprule
      \multirow{2}{*}{Model} & Total   & \multicolumn{3}{c||}{All} & \multicolumn{3}{c|}{Assertion} & \multicolumn{3}{c|}{Timeout} & \multicolumn{3}{c}{Exception} \\   
                            & Studied & $@1$                      & $@3$                           & $@5$                         & $@1$                           &  $@3$         & $@5$         & $@1$ & $@3$ & $@5$ & $@1$        & $@3$        & $@5$        \\ \midrule
      EM(F)                  & 357     & 77                        & 139                            & 163                          & \textbf{100}                   &  \textbf{163} & \textbf{179} & 21   & 30   & 38   & 53          & 90          & 107         \\
      PM$^*$(F)              & 357     & 82                        & 151                            & 183                          & \textbf{108}                   &  \textbf{185} & \textbf{206} & 23   & 35   & 44   & 60          & 97          & 117         \\
      PM$^+$(F)              & 357     & 86                        & 155                            & 186                          & \textbf{114}                   &  \textbf{183} & \textbf{206} & 22   & 36   & 45   & 61          & 99          & 119         \\
      LR(F)                  & 357     & 90                        & 152                            & 186                          & \textbf{118}                   &  \textbf{181} & \textbf{213} & 40   & 79   & 88   & 66          & 109         & 131         \\
      MLP(F)                 & 357     & 85                        & 150                            & 180                          & \textbf{121}                   &  \textbf{189} & \textbf{210}          & 43   & 76   & 91   & 60          & 106         & 129         \\ \midrule
      EM(F+P)                & 224     & \textbf{84}                        & \textbf{108}                   & \textbf{114}                          & 72                             &  89           & 96           & 7    & 11   & 16   & 55          & 64          & 68          \\
      PM$^*$(F+P)            & 224     & 49                        & 74                             & 86                           & \textbf{50}                    &  \textbf{76}  & \textbf{90}  & 23   & 42   & 57   & 49          & 71          & 80          \\
      PM$^+$(F+P)            & 224     & 33                        & 57                             & \textbf{68}                  & \textbf{34}                    &  57           & \textbf{68}  & 20   & 34   & 50   & \textbf{34} & \textbf{59} & \textbf{68} \\
      LR(F+P)                & 224     & 88                        & \textbf{120}                   & \textbf{129}                 & \textbf{89}                    &  117          & 128          & 23   & 41   & 51   & 77          & 104         & 115         \\
      MLP(F+P)               & 224     & \textbf{113}              & 145                            & 159                          & 112                            &  \textbf{147} & \textbf{160}          & 29   & 50   & 55   & 91          & 120         & 135         \\
      PC(F+P) & 224 & 96 & 143 & 157 & \textbf{100} & \textbf{145} & \textbf{160} & 22 & 36 & 43 & 78 & 113 & 121 \\
      \bottomrule
    \end{tabular}
  }
\end{table*}
A mutated program can cause a test failure due to many different reasons, such 
as assertion (i.e., test oracle) violation, uncaught exception, or timeout. 
All these reasons are normally marked as a kill. While all three reasons do
reveal some dependency between the mutated location and the test outcome
(otherwise the mutant would not be killed), we suspect that different kill
reasons may have varying degrees of importance for fault localisation. 
Assertion violations would imply that the test oracles actually capture
the correct program behaviour. Uncaught exceptions and timeouts, however,
may only show coincidental impacts of the mutation.

Considering the relative importance of different kill reasons, we 
investigate whether filtering out the kill matrix based on the exact reason of 
test failure has any impact on the localisation effectiveness. This is
partly motivated by the use of failure messages by 
TraPT~\cite{li2017transforming}.
We train \name models using one of three kill reasons,
and compare their results to those of models trained
using all three reasons. Kill reasons supported by Major are: assertion 
violations (``Assertion"), timeouts (``Timeout"), and uncaught exceptions (``Exception").

Table~\ref{tab:kill_reason_filtering} shows the results of $acc@n$ metrics for
\name models of three different kill reasons. For all F models, using only mutants 
killed due to the assertion failures shows the best performance in terms 
of $acc@1$ and $acc@3$, adding support to our assumption that
assertion violations reflect test oracles of correct program behaviour better than others.
Timeouts appear to be the weakest signal.


However, for F+P models, the unfiltered original results (``All") often show 
the best performance. This trend reveals a seemingly 
counter-intuitive, yet fundamental intuition about \name: test cases in $\mathbf{T}_f$ 
and $\mathbf{T}_p$ contribute to localisation in different ways. If a test case $t$ is in $\mathbf{T}_f$, all mutants 
killed by $t$ earlier suggest that their locations may contain the fault. 
However, if $t \in \mathbf{T}_p$, all mutants killed by $t$ earlier suggests that their 
locations may \emph{not} contain the fault that is detected by $t' \in \mathbf{T}_f$.
Consequently, kill reason filtering can make the contributions from tests in 
$\mathbf{T}_f$ more precise (i.e., to only reflect real fault detection), but may also 
reduce the total amount of contributions from tests in $\mathbf{T}_p$ because it removes 
potential locations that could have been \emph{excluded} by being associated 
with a test in $\mathbf{T}_p$. This explains why, for F+P models, using only Assertion as
the kill reason cannot dominate the results. Note that the distribution of 
kills between Assertion, Timeout, and Exception is likely not uniform, which we
also think contributes to the mixed results of F+P models, combined with 
program semantics.




\section{Threats to Validity}
\label{sec:threats}

Given the controlled setting for our experiments and the clearly defined
objective measures, there are few threats to the  validity of our study.
There are some threats to internal validity that are inherent to any mutation
analysis and hard to completely avoid, such as non-determinism caused by
mutation and equivalent mutants, which have been discussed in
Section~\ref{sec:result}. Similarly, we see few threats to the construct and
conclusion validity. The metrics we used are standard in the fault localisation
literature. Additionally, we acknowledge that the small number of samples
(sometimes less than 10) used in the Mann-Whitney U test may be a potential
threat to the statistical power of our results.

Another potential threat to validity can arise from the offline costs of
\name. As \name has to be ready for diverse future faults, it needs a larger
kill matrix than other MBFL techniques that only need a partial kill matrix
relevant to the fault. Therefore, if we compare these techniques within a single
debugging scenario, \name may be more expensive. Assuming that the cost of
running the test suite is the same for all mutants, then we can estimate the
offline costs by comparing the number of mutants that need to be executed
against the test suite. In our analysis on the subject programs, the ratio
between the mutants that are relevant to a single fault on average and the total
number of mutants is approximately 1 to 6. This means that after localising
about 6 faults, performing a full-scale mutation analysis in advance can result
in a more usable MBFL technique than analyzing mutations for each test failure,
as it allows for parallelisation and offline execution. Additionally, reducing
the offline costs of \name can be achieved by creating a partial kill matrix for
program elements more likely to have faults based on defect prediction
strategies.

As suggested by Steimann et al.~\cite{Steimann:2013sf}, the presence of multiple
faults can be the threat to validity. This can complicate the fault localization process and lead to less
accurate results. In practice, it is not always possible to achieve perfect bug
detection, resulting in increased effort required to detect faults.
Additionally, we should note that our assumption that a failing test case
executes at least one faulty position in the source code may introduce a
potential bias and influence the validity of our results.

Rather, the main threat of our study is to its external validity. Even though we
studied five different subjects from the real-world \dfj benchmark to mitigate
this threat, this does not allow us to generalise to many, other programs and
test suite contexts. Still, there was enough variation among the subjects for us
to identify \name's dependence on the granularity of the test cases.

\section{Related Work}
\label{sec:related_work}

\subsection{Mutation-based Fault Localisation}
A number of MBFL techniques have been proposed
in the literature. Metallaxis uses SBFL-like formulas to measure the similarity 
between failure patterns of the actual fault and 
mutants~\cite{Papadakis:2015sf,Papadakis:2012fk}. MUSE~\cite{Moon:2014ly}, 
and its variation MUSEUM~\cite{Hong:2017qy}, depend on two principles: first, 
if we mutate already faulty parts of the program, it is unlikely that we will 
observe more failing test cases, and we may even observe partial fixes, and 
second, if we mutate non-faulty parts, tests that used to pass are now likely
to fail. MUSE and MUSEUM define their suspiciousness scores using the
ratios of fail-become-pass and pass-become-fail tests. 
TraPT is similar to MUSE and MUSEUM in nature, but transforms both the
output messages of failing tests, to distinguish different types of exceptions,
and the test code itself, to prevent early program termination due to the 
assertion violation that precludes collecting information of other assertions
~\cite{li2017transforming}. 

MBFL techniques suffer from the cost of mutation analysis, as it requires 
a massive number of test executions against the generated mutants. To reduce 
the number of mutants needed to be inspected, Metallaxis~\cite{Papadakis:2015sf}
adopts mutant sampling while HOTFUZ~\cite{jang2021hotfuz} combines first-order 
mutants to build higher-order mutants and computes suspicious score only on them.
Their results showed that this reduces the number of 
total mutants as well as equivalent mutants that contribute nothing to the 
fault localisation. Despite such efforts to reduce the costs, all existing MBFL 
techniques inevitably mutate the faulty program and execute tests once testing 
is finished. In contrast, \name allows the mutation analysis to be performed 
ahead of time.


\subsection{Deep Learning Based Fault Localisation}

DeepFL~\cite{li2019deepfl} is a Deep Learning (DL) based learning-to-rank
FL technique that integrates the results of multiple FL techniques. By
aggregating a variety of information such as program spectrum, mutation, and
source code metrics, they achieve better FL performance than the individual
techniques. On the other hand, DEEPRL4FL~\cite{li2021fault} leverages a coverage
matrix, a data flow graph, and abstract syntax trees to learn to localise the
faults, using Convolutional Neural Networks (CNNs). Both techniques have shown
promising results through utilising multiple sources of information. In
contrast, \name stands out as a single MBFL technique that is orthogonal to the
existing MBFL techniques, presenting a potential opportunity for integration as
a distinctive feature in the synthesis of FL results.

\subsection{Bayesian Inference for Fault Localisation}
\name was initially formulated based on Bayesian analysis to infer likely fault 
locations given test information. In the context of fault localisation, 
Abreau et al.~\cite{Abreu:2009qy} have introduced \textsc{Barinel}, a SBFL technique that 
adopts Bayesian reasoning to generate candidate sets of multiple fault
locations. To the best of our knowledge, \name is the first MBFL technique
that uses Bayesian inference as well as other statistical inference techniques.
While \name also uses dynamic information from mutation, 
the mutation analysis can be performed ahead of time, which allows the cost 
to be amortised over multiple development iterations, and provides faster 
feedback.

\subsection{Mutant Subsumption}
The subsumption relationship between two mutants appears when both are
killed by the tests and all tests that kill the one also kill the other. Therefore,
this indicates a redundancy of the subsumed mutant~\cite{kurtz2014mutant},
allowing the mutation score to be inflated. As a result, building a minimal set
of mutants can enhance the validity and efficacy of mutation
analysis~\cite{ammann2014establishing}. We showed that \name can exploit this by
removing the subsumed mutants in the kill matrix and achieve considerably better
results than the normal setting using all mutants.

\subsection{Predictive Mutation Testing}
 A recent advance to deal with the high costs of mutation testing is
Predictive Mutation Testing (PMT)~\cite{zhang2018predictive, mao2019extensive}.
PMT predicts the mutation score as well as test results of each mutant. Since
PMT uses historical mutation testing results as its features to train the
predictive model, PMT and \name have a similar scenario that amortises the cost
of mutation testing. However, PMT is originally designed to predict whether a
test suite can kill the mutants, so it is hard to apply to \name that requires
the test-case level results of mutant kills. \se~\cite{Kim2021ax} provides a
test-case level prediction of mutant kills using the tokens in the source code
and test as well as the dynamic information of the mutants. The finer-grained
prediction of \se enables its combined use with \name, which helps \name to
deal with newly introduced test cases by simply predicting which mutants they
can kill.

\section{Conclusion}
\label{sec:conclusion}
 In this paper, we introduce \name, a Mutation Based Fault Localisation
  (MBFL) technique utilising mutation analysis results of the earlier version.
  \name models the relations between the location of the artificial faults
  (i.e., mutants) and their test results via several modelling schemes such as
  statistical inference or machine learning techniques. After the faults are
  observed, \name infers their locations without any mutant executions. We
  evaluate \name on the real-faults in \dfj dataset and the results show that \name can locate 113 faults on
  the first rank out of 224 faults. Even if we simulate \name on the predicted kill
  matrix, \name still localises 95 faults on the first rank out of 194 faults,
  supporting its practical applicability. Moreover, we demonstrate that removing
  the subsumed mutants significantly improves the localisation accuracy of \name.

\bibliography{newref}

\end{document}